\documentclass{fldauth}
\pdfoutput=1
\usepackage{subfigure}
\usepackage{amssymb}

\begin{document}
\FLD{1}{6}{00}{28}{00}

\runningheads{M. Robinson and J. Monaghan}
{DNS of decaying 2D turbulence using SPH}

\title{Direct Numerical Simulation of decaying two-dimensional turbulence in a no-slip square box using Smoothed Particle Hydrodynamics}

\author{Dr. Martin Robinson\affil{1}\corrauth, Prof. Joseph J. Monaghan \affil{1}}

\address{\affilnum{1}\ School of Mathematical Sciences\\ 
Monash University\\
Vic 3800 Australia}

\corraddr{m.j.robinson@ctw.utwente.nl or University of Twente, Faculty of Engineering, Postbus 217, 7500 AE Enschede, The Netherlands}

\noreceived{}
\norevised{}
\noaccepted{}

\begin{abstract}

This paper explores the application of SPH to a Direct Numerical
Simulation (DNS) of decaying turbulence in a two-dimensional no-slip
wall-bounded domain. In this bounded domain, the inverse energy cascade, and a
net torque exerted by the boundary, result in a spontaneous spin up of the
fluid, leading to a typical end state of a large monopole vortex that fills the
domain. The SPH simulations were compared against published results using a
high accuracy pseudo-spectral code. Ensemble averages of the kinetic energy,
enstrophy and average vortex wavenumber compared well against the
pseudo-spectral results, as did the evolution of the total angular momentum of
the fluid. However, while the pseudo-spectral results emphasised the importance
of the no-slip boundaries as generators of long lived coherent vortices in the
flow, no such generation was seen in the SPH results. Vorticity filaments
produced at the boundary were always dissipated by the flow shortly after
separating from the boundary layer. The kinetic energy spectrum of the SPH
results was calculated using a SPH Fourier transform that operates directly on
the disordered particles. The ensemble kinetic energy spectrum showed the
expected $k^{-3}$ scaling over most of the inertial range. However, the
spectrum flattened at smaller length scales (initially less than 7.5 particle spacings
and growing in size over time), indicating an excess
of small-scale kinetic energy.

\end{abstract}


\section{Introduction}

The purpose of this research is to provide information on how well SPH models
turbulent flow without the additional complication of a turbulence model. While
there has been many SPH turbulence models presented in the literature, these
 have not shown a clear advantage over standard SPH, and there is
some speculation that the method naturally contains an artificial LES-type
model \cite{cleary93boundary,ting06simulation}. It is therefore
important to explore how well standard SPH can simulate various classes of
turbulent flow, in order to determine the suitability of the method to turbulence
applications and to provide results that can inform the development of any
future SPH turbulence models. 

\subsection{SPH Turbulence}

In the past decade there have been numerous turbulence models proposed for SPH.
The majority of these models have focused on using the LES method combined with a
Smagorinsky model for the sub-grid (or sub-particle in this case) scale
\cite{violeau07numerical,dalrymple06numerical,shao06simulation,issa05firstAttempt,shao04simulating}.
However, there have also been proposals based on the standard one and
two-equation turbulent transport equations \cite{violeau07numerical}, the
Renolds Stress model \cite{violeau07numerical}, stochastic pdf models
\cite{violeau02twoAttempts,welton98twodimensional} and a LANS-$\alpha$ model
\cite{monaghan02SPHturbulence,monaghan04energy}.

Ting et al. \cite{ting06simulation} used an SPH LES model to simulate the turbulent flow
over a backwards facing step. Ting et
al. found that standard SPH produced reasonable results and that the
addition of the turbulence models did not improve these significantly. They
echoed the conclusions by Cleary and Monaghan \cite{cleary93boundary} that SPH can be considered to
have a form of LES already built in and that SPH involves some sort of
dissipative term that prevents the accumulation of energy at the sub-grid
scales. 

All of these papers involve the addition of a turbulence model to standard SPH.
However, there has been few attempts to study how well standard SPH can model
the full range of turbulent scales in a Direct Numerical Simulation (DNS).
Mansour \cite{mansour07SPH} has performed the only SPH DNS prior to this paper. He used
standard SPH as well as the LANS-$\alpha$ based model proposed by Monaghan
\cite{monaghan02SPHturbulence} (termed $\alpha$-SPH) in order to simulate
forced 2D incompressible turbulence in a periodic box. Mansour found that while
SPH reproduces an inverse energy cascade, its strength is much weaker than
expected. This was attributed to the weakness of the SPH viscosity term at
small scales and the resultant action of this term over a much broader range of
scales than expected by theory. The $\alpha$-SPH model failed to improve on
these results. In fact, the model caused an increase in numerical kinetic
energy at short scales, although Mansour notes that a further increase in the
length scale of the smoothing should reverse this trend. This was only
attempted in 1D simulations due to the computational requirements of the model.

\subsection{Wall-bounded Two-Dimensional Turbulence}

We have chosen to simulate turbulence in a wall-bounded domain, which will give
a clearer picture of how SPH turbulence behaves in a more practical setting
than the traditional periodic box. Another motivation for this choice is the
uncertainty over how well a periodic box approximates an infinite domain. Much
of turbulence theory assumes that the turbulence is isotropic and homogeneous
and acts in an infinite domain. A periodic box is normally assumed to
approximate an infinite domain for scales below the size of the box.
Large-scale dissipation terms are usually used to prevent energy accumulating
at the box length scale, but Tran and Bowman \cite{tran03dual} argue
that a large-scale dissipation term would also remove a significant proportion
of the enstrophy and that any energy not removed would be reflected down to the
direct enstrophy range, altering the results from the unbounded case.  Davidson
\cite{davidson04turbulence} notes that the periodic boundaries cause artificial
anisotropy and long-range correlations in the flow at scales near the size of
the box. Lowe \cite{lowe01DNS2DTurbPeriodic} investigated the effects of
periodic boundary conditions in two-dimensional turbulence and found that they
were felt at length scales 100 times smaller than the size of the box. Perhaps
the most practical problem with turbulence in a periodic box is that it ignores
the effects that non-slip and stress-free boundary conditions can have on the
turbulence. This is a natural concern for any real-world turbulent flow.

The reasons for simulating two-dimensional turbulence are largely pragmatic.
While it is often argued that two-dimensional turbulence is not useful due to
its fundamental differences from three-dimensional turbulence, it nevertheless
makes an ideal benchmark test for any numerical method. Turbulence has a
well-deserved reputation of being difficult to simulate, due to the complex,
chaotic nature of the flow as well as the high resolution requirements.
Two-dimensional turbulence reduces the resolution requirements significantly.
In addition to the reduction in dimensionality, the smallest length scale in
the flow $\eta$ scales much slower with the Reynolds Number ($\eta \propto Re^{-\frac{1}{2}}$
rather than $\eta \propto Re^{-\frac{3}{4}}$). At the same time, the move from
three dimensions to two surprisingly makes the flow more difficult to simulate
correctly. There is a much closer link between the small and large length
scales in two dimensions, due to the inverse energy cascade. If these small
scales are not simulated correctly, it could disrupt the cascade of energy to
lower wavenumbers and hence change the large-scale properties of the flow.
Conversely, if the small scales are incorrectly modelled in three dimensions,
the results will only affect these and smaller scales.

Coupled with the difficulty of the simulation, there is a wide variety of
numerical, experimental and theoretical results to compare against. One of the
first numerical simulations of turbulence was performed by Lilly
\cite{lilly69numerical}. Going back even further, the first major experiment on
turbulence is usually attributed to Reynolds \cite{reynolds83experimental}.
Since this time there has been no shortage of subsequent experimental and
numerical results. There is also a large body of theoretical predications
thanks to the pioneering work of Richardson \cite{richardson26atmospheric},
Kolmogorov \cite{kolmogorov41decay}, Batchelor
\cite{batchelor69energySpectrum}, Kraichnan \cite{kraichnan67inertialRange} and
many others since. 

\section{Decaying Wall-bounded Two-Dimensional Turbulence}

Two-dimensional turbulence retains the same properties of randomness and
chaotic advection that are the hallmarks of its three-dimensional equivalent.
However, the reduction in dimensionality has profound effects on the main
turbulent processes. The origin of many of these effects lies in the lack of
vortex stretching in two dimensions, meaning that the only change in vorticity
$\omega$ (a scalar in 2D) is due to viscous forces. In a high $Re$ flow these will be minimal
and hence the vorticity is materially conserved as $Re\rightarrow0$. The
primary action of the turbulence on a volume of fluid is to stretch it out into
a filamentary structure. Since the vorticity of a material point is relatively
constant compared with the turbulent time scale, an initial vorticity element
will also be stretched out over time into a filament.  This will continue until
the vorticity gradients become too great and the vorticity filament is
destroyed due to viscous forces. 

This process is given the term \emph{direct enstrophy cascade}, where enstrophy
is defined as $\omega^2/2$. It refers to the movement of enstrophy from the
large to small length scales until it is mopped up at the dissipation scale.
The word cascade implies that this action is local in wavenumber space (i.e. only flow structure of similar sizes interact) and that
the enstrophy moves through all scales on the way to the bottom. However, vortices of quite
different length scales can interact in the flow \cite{davidson04turbulence},
indicating that this is not the case.  Nevertheless, the term cascade is well
established in the literature.

This leads to the hallmark of two-dimensional turbulence, the \emph{inverse
energy cascade}. Rather than the kinetic energy held in large scale vortices
moving to smaller and smaller scales, as occurs in three dimensions, the
opposite is true for two-dimensional turbulence. The process by which this
occurs is not quite as clear as the enstrophy cascade.  Proposals include the
merger of like-signed vortices \cite{ayrton19,fujiwhara21} and that
the growth in kinetic energy length scales is related to the increasing length
of the vorticity filaments \cite{davidson04turbulence}.

In a non-infinite domain, the inverse energy cascade will result in energy
accumulating in the largest mode corresponding to the size of the domain.
Kraichnan \cite{kraichnan67inertialRange} predicted this and compared the process to
Einstein-Bose condensation. For decaying turbulence in a periodic box the end
steady-state condition will be a pair of vortices of opposite sign. For an
example, see the numerical results by Smith and Yakhot \cite{smith93bose}. For this paper, we
are interested in the simulation of wall-bounded turbulence.
Clercx et al. \cite{clercx99Decaying2DTurbinSquareCont} have performed a pseudo-spectral
simulation of decaying turbulence in a two-dimensional square box using no-slip
boundaries. He found that in the majority of cases pressure forces
perpendicular to the wall impart a net torque to the flow. As the turbulence
decays, this torque combines with the inverse energy cascade to produce a
single large monopole vortex that fills the entire domain. However, this end
state was dependent on the initial turbulent velocity field, which was randomly
generated for each simulation. In a small number of cases the net torque would
be very small or non-existent, in which case the turbulence would decay into a
pair of vortices of opposite sign. Clercx et al. also found that the boundaries
were a significant source of vorticity and vorticity gradients in the flow.
Whenever a vortex came near a boundary it would create a boundary layer
response. This boundary layer could separate from the wall and move into the
interior of the flow as a high intensity vorticity filament. There filaments
would roll up to form coherent vortices that would persist for long times in
the flow. 

Li et al. \cite{li96decaying,li97twoDimensional} have also performed decaying turbulence
simulations in a circular wall-bounded domain. They too note the importance of
the boundaries in injecting vorticity into the flow, as well as the significant
increase in kinetic energy decay this causes as compared with the periodic
case. 

We have setup an SPH decaying turbulence simulation with an identical geometry
and initial velocity field as that presented in the pseudo-spectral results by
Clercx et al. \cite{clercx99Decaying2DTurbinSquareCont}. This will allow the comparison of
the SPH results with these highly accurate simulations.

\section{Smoothed Particle Hydrodynamics}\label{Chap:theory}

Smoothed Particle Hydrodynamics \cite{gingold77smoothed, lucy77numerical,
monaghan05SPH} is a Lagrangian scheme, whereby the fluid is discretised into
particles that move with the fluid velocity. Each particle is assigned a mass
and can be thought of as the same volume of fluid over time. The fluid
variables and the equations of fluid dynamics are interpolated over each
particle and its nearest neighbours using a Gaussian-like kernel with compact
support. 

SPH is based on the idea of kernel interpolation. A fluid variable
$A(\mathbf{r})$ (such as velocity or density) is interpolated using a kernel
$W$, which depends on the smoothing length variable $h$.

\begin{equation}\label{Eq:integralInterpolant}
A(\mathbf{r}) = \int A(\mathbf{r'})W(\mathbf{r}-\mathbf{r'},h)d\mathbf{r'}.
\end{equation}

To apply this to the discrete SPH particles, the integral is replaced by a sum
over all particles, commonly known as the \emph{summation interpolant}. To
estimate the value of the function $A$ at the location of particle $a$ (denoted
as $A_a$), the summation interpolant becomes

\begin{equation}\label{Eq:summationInterpolant}
A_a = \sum_b m_b \frac{A_b}{\rho_b} W_{ab}.
\end{equation}

where $m_b$ and $\rho_b$ are the mass and density of particle $b$.  The volume
element $d\mathbf{r'}$ of Equation \ref{Eq:integralInterpolant} has been
replaced by the volume of particle $b$ (approximated by $\frac{m_b}{\rho_b}$).
This is the normal trapezoidal quadrature rule.  The kernel function is denoted
by $W_{ab} = W(\mathbf{r}_a-\mathbf{r}_b,h)$. The dependence of the kernel on
the smoothing length $h$ and the difference in particle positions is not
explicitly stated for readability.

The spatial derivative of (\ref{Eq:integralInterpolant}) is found by writing the
derivative as \cite{monaghan05SPH}

\begin{equation} \label{Eq:derivativeSymettric}
\nabla A_a = \frac{1}{\Phi} \left ( \nabla (\Phi A) - A \nabla \Phi \right ),
\end{equation}

where $\Phi$ is any differentiable function. This ensures that the final SPH
equation is symmetric between particle pairs.  When this function is converted
to SPH form, it becomes

\begin{equation} \label{Eq:derivativeInterpolant}
\nabla A_a = \frac{1}{\Phi_a} \sum_b m_b \frac{\Phi_b}{\rho_b} (A_b-A_a) \nabla W_{ab}.
\end{equation}

The kernel most commonly used in SPH calculations is the Cubic Spline. This is
a third-order polynomial with compact support that is based on the family of
spline functions in \cite{schoenberg46contributions}. The Cubic Spline kernel
is defined as 

\begin{eqnarray}\label{Eq:CubicSplineKernel}
W(s) = \frac{\beta}{h^d}
\begin{cases}
(2-q)^3 - 4(1-q)^3                         & \text{for } 0 \leq q < 1, \\
(2-q)^3                                    & \text{for } 1 \leq q < 2, \\ 
0                                          & \text{for } q > 2.
\end{cases}
\end{eqnarray}

where $q=r/h$, $d$ is the dimensionality of the kernel and $\beta$ is a constant equal to
$1/6$, $5/(14\pi)$ and $1/(4\pi)$ for one, two and three dimensions respectively.

The rate of change of density, or continuity equation, is given by

\begin{equation}
\frac{D\rho}{Dt} = -\rho \nabla \cdot \mathbf{v}.
\end{equation}

Using (\ref{Eq:derivativeInterpolant}) to estimate $\nabla \cdot \mathbf{v}$ with $\Phi=\rho$, this becomes

\begin{equation} \label{Eq:changeInDensity}
\frac{D\rho_a}{Dt} = \sum_b m_b \mathbf{v}_{ab} \cdot \nabla_a W_{ab},
\end{equation}

where $\mathbf{v}_{ab}=\mathbf{v}_a-\mathbf{v}_b$. 

SPH originated from the astrophysical community \cite{gingold77smoothed,
lucy77numerical}, where it was applied to compressible gas. For incompressible
flows, SPH algorithms usually use a quasi-compressible formulation, where the
density varies by less than 1\% between particles. 

The equation of state commonly used in the quasi-compressible SPH literature is
(originally defined by Cole in \cite{cole48underwater})

\begin{equation}
P = B \left ( \left ( \frac{\rho}{\rho_0} \right )^\gamma - 1 \right ),
\end{equation}

where $\gamma = 7$ is a typical value and $\rho_0$ is a reference density that
is normally set to the density of the fluid. 

At the reference density ($\rho = \rho_0$) the constant $B$ can related to the
speed of sound $c_s$ by

\begin{equation}
c_s^2 =  \left. \frac{\partial P}{\partial \rho} \right |_{\rho=\rho_0} = \frac{\gamma B}{\rho_0}.
\end{equation}

The speed of sound, and hence $B$, is chosen to keep the density variation
between particles small (usually less than 1\%). Since

\begin{equation}
\frac{| \delta \rho |}{\rho} = \frac{v^2}{c_s^2},
\end{equation}

in order to keep $\frac{| \delta \rho |}{\rho} < 0.01$, the constant $B$ must be set to

\begin{equation}
B \ge \frac{100\rho_0 v_m^2}{\gamma},
\end{equation}

where $v_m$ is an estimate of the maximum velocity of the flow. 

Neglecting the viscous term, the momentum equation depends only on the pressure gradient

\begin{equation}
\frac{D\mathbf{v}}{Dt} = \frac{1}{\rho} \nabla P.
\end{equation}

Recall the symmetric form of the derivative introduced in (\ref{Eq:derivativeSymettric}). Taking $\Phi=1/\rho$ this becomes

\begin{equation} 
\frac{1}{\rho} \nabla P = \nabla \left ( \frac{P}{\rho} \right ) + \frac{P}{\rho^2} \nabla \rho.
\end{equation}

Using the SPH summation form of the spatial gradient (Equation \ref{Eq:derivativeSymettric}) this becomes

\begin{equation}\label{Eq:sphJustPressureForce}
\frac{D\mathbf{v_a}}{Dt} = -\sum_b m_b \left ( \frac{P_b}{\rho_b^2} + \frac{P_b}{\rho_b^2} \right ) \nabla_a W_{ab}.
\end{equation}

Viscosity is included by adding a viscous term $\Pi$ to the SPH momentum equation

\begin{equation} \label{Eq:momentumWithVisc}
\frac{D\mathbf{v_a}}{Dt} = -\sum_b m_b \left ( \frac{P_b}{\rho_b^2} + \frac{P_b}{\rho_b^2} + \Pi_{ab} \right ) \nabla_a W_{ab}.
\end{equation}

The SPH literature contains many different forms for $\Pi$. We have used the term proposed by 
Monaghan \cite{monaghan97SPHRiemannSolvers}, which is based on the dissipative
term in shock solutions based on Riemann solvers. This viscosity was
initially derived to prevent the unphysical penetration of colliding gas clouds
but is also successful when modelling the viscosity for incompressible fluids.
For this viscosity

\begin{equation}\label{Eq:monaghansViscousTerm}
\Pi_{ab} = - \alpha \frac{v_{sig} (\mathbf{v}_{ab} \cdot \mathbf{r}_{ab} )}{2 \overline{\rho}_{ab} |\mathbf{r}_{ab}|},
\end{equation}

where $v_{sig} = 2(c_s + |\mathbf{v}_{ab} \cdot \mathbf{r}_{ab}| /
|\mathbf{r}_{ab}| )$ is a signal velocity that represents the speed at which
information propagates between the particles. The constant $\alpha$ gives the
viscosity strength and can be related to the dynamic viscosity
$\mu$ using \cite{monaghan05SPH}

\begin{equation}
\mu = \rho \alpha h c / S.
\end{equation}

where, for the Cubic Spline kernel, $S=112/15$ for two dimensions and $S=10$ for three.

The particle's position and velocity were integrated using the
Leapfrog second order method. This is a geometric, or symplectic, integrator.
This class of integrators (See \cite{hairer03geometric} for more details) are commonly used for
molecular and celestial mechanics simulations. While they lack the absolute
accuracy of methods like the Runge-Kutta, they are designed to preserve
(exactly) the Lagrangian of a system and so have much improved long-term
behaviour. They are also reversible in time (in the absence of viscosity). 
To preserve the reversibility of the simulation, $d\rho/dt$ was
calculated using the particle's position and velocity at the end of the
timestep, rather than the middle as is commonly done. The full integration scheme is given by 

\begin{align}
\mathbf{r}^{\frac{1}{2}} &= \mathbf{r}^{0} + \frac{\delta t}{2} \mathbf{v}^{0}, \\
\mathbf{v}^{\frac{1}{2}} &= \mathbf{v}^{0} + \frac{\delta t}{2} F(\mathbf{r}^{-\frac{1}{2}},\mathbf{v}^{-\frac{1}{2}},\rho^{-\frac{1}{2}}), \\
\rho^{\frac{1}{2}} &= \rho^{0} + \frac{\delta t}{2} D(\mathbf{r}^0,\mathbf{v}^0), \label{Eq:timestepDensity1} \\
\mathbf{v}^{1} &= \mathbf{v}^{0} + \delta t F(\mathbf{r}^{\frac{1}{2}},\mathbf{v}^{\frac{1}{2}},\rho^{\frac{1}{2}}), \\
\mathbf{r}^{1} &= \mathbf{r}^{\frac{1}{2}} + \frac{\delta t}{2} \mathbf{v}^{1}, \\
\rho^{1} &= \rho^{\frac{1}{2}} + \frac{\delta t}{2} D(\mathbf{r}^1,\mathbf{v}^1), \label{Eq:timestepDensity2}
\end{align}

where $\mathbf{r}^0$, $\mathbf{r}^{1/2}$ and $\mathbf{r}^1$ is $\mathbf{r}$ at
the start, mid-point and end of the timestep respectively. The timestep $\delta
t$ is bounded by the standard Courant condition

\begin{equation}
\delta t_1 \le \min_a \left ( 0.8 \frac{h_a}{v_{sig}} \right ),
\end{equation}

where the minimum is taken over all the particles. 

The no-slip boundaries in the simulation were modelled using four layers of
immovable SPH particles. These boundary particles were identical to the other
fluid particles in every way except that their positions and velocities were
constant. That is, only Equations \ref{Eq:timestepDensity1} and
\ref{Eq:timestepDensity2} of the timestepping scheme are applied to the
boundary particles.

\section{Initial Turbulent Velocity Field}

The initial velocity field was chosen to match that used by Clercx et al.
\cite{clercx99Decaying2DTurbinSquareCont}. The SPH particles were positioned on
a grid covering a square domain defined by $-1 \le x \le 1$ and $-1 \le y \le
1$. Each particle's velocity was calculated from a 2D 65x65 Chebyshev series.

\begin{equation}\label{Eq:decayTurb2Dcheb}
v_a = \sum_{n=0}^{65} \sum_{m=0}^{65} C_{nm} T_n(x_a) T_n(y_a),
\end{equation}

where $x_a$ and $y_a$ are the x and y coordinates of particle $a$'s position and 
$T_n(x) = \cos \left (n \cos^{-1}(x) \right )$ is the $n^{th}$ Chebyshev
polynomial. The coefficients $C_{nm}$ were randomly generated from a zero-mean
Gaussian distribution with variance $\sigma_{nm}$ given by

\begin{equation}
\sigma_{nm} = \frac{n}{\left [ 1+\left (\frac{1}{8}n \right )^4 \right ]} \frac{m}{\left [ 1+\left (\frac{1}{8}m \right )^4 \right ]}.
\end{equation}

\begin{figure}[htbp] 
\begin{center}
\includegraphics[width=0.6\textwidth]{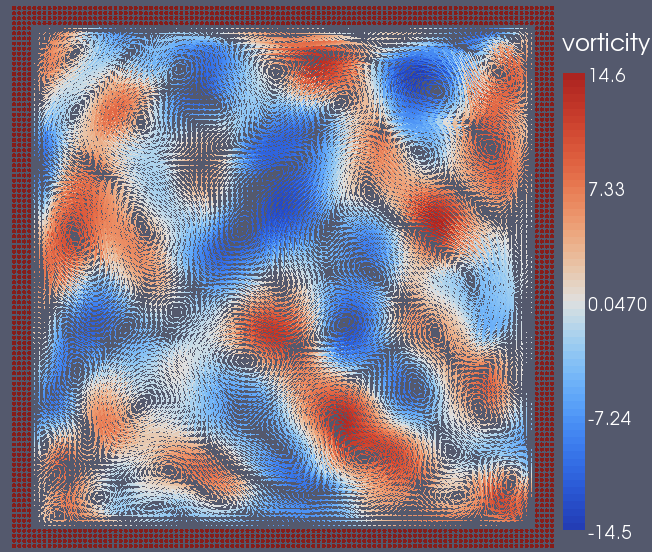} 
\caption{An typical initial velocity field for the decaying turbulence simulation. Arrows
depict the velocity field and are coloured by the vorticity. The four layers of red dots around the edges of the box 
denote boundary SPH particles}\label{Fig:initialConditions} 
\end{center} 
\end{figure}

In order to ensure a quiet start at the no-slip boundaries, the velocity field
was gradually reduced down to zero near the boundaries by multiplying the velocity with
a function $f(x)f(y)$ where

\begin{equation}\label{Eq:decayTurbrolloff}
f(x) = 1 - \exp(-100(1-x^2)^2).
\end{equation}

The resultant velocity field is not necessarily divergence-free. However, the
pseudo-spectral code described by Clercx et al. ensures that it is by projecting
the velocity onto a divergence-free subspace. Following this example, we have
performed a similar operation on velocity field defined by
(\ref{Eq:decayTurb2Dcheb}) - (\ref{Eq:decayTurbrolloff}).  

Any velocity field can be separated into its divergence-free component
$\mathbf{v}_d$ and the gradient of a scalar $\phi$

\begin{equation}
\mathbf{v} = \mathbf{v}_d + \nabla \phi.
\end{equation}

Taking the divergence of both sides gives

\begin{equation}
\nabla \cdot \mathbf{v} = \nabla^2 \phi.
\end{equation}

Since the SPH particles are initially on a grid, this equation can be solved
for $\phi$ using a second order finite difference method. The divergence free
velocity field $\mathbf{v}_d = \mathbf{v}-\nabla \phi$ is then normalised so
that the total kinetic energy $E(0)=1$ per unit mass. Figure
\ref{Fig:initialConditions} shows an example of an initial velocity field
generated using this method. 

In order to be absolutely consistent, the pressure of the particles should be
set to match the given velocity field. However, this was not expected to alter
the results significantly as the SPH pressures quickly evolve to match the velocity
field over a time scale set by the sound speed, which is 10 times greater than
the maximum velocity of the flow.

\section{Simulation Parameters}

The Reynolds Number was set to $Re=\ell U/\nu=1500$ based on the half-width of the
box $\ell = 1$ and the initial RMS velocity $U=1$ of the particles. The simulation
time $t$ is scaled by $\ell /U$ and is dimensionless. $t=1$ is comparable to one
eddy turn-over time \cite{clercx99Decaying2DTurbinSquareCont}.

The particle densities and the reference density were set to $\rho = \rho_0 =
1000$. Therefore, the initial pressure field of the particles will be constant and equal to zero.

The simulated domain was a square box defined by $(-1 \le x,y \le 1)$. The
number of particles in the box was set to $300 \times 300$ and the ratio
between smoothing length and particle separation was $h/\Delta p = 1.95$.
Convergence studies (shown in Section \ref{Sec:ConvergenceStudy}) have
indicated that the results of the decaying turbulence simulations are
sufficiently converged using these resolution parameters.

\section{Vorticity Evolution}

The vorticity for each particle $a$ is determined as follows. A linear estimate
of the velocity field around particle $a$ is defined as

\begin{align}
v_x(x,y) &= a_{1,1} (x-x_a) + a_{1,2} (y-y_a) \\ 
v_y(x,y) &= a_{2,1} (x-x_a) + a_{2,2} (y-y_a),
\end{align}

where $\mathbf{v} = (v_x,v_y)$ is the linear velocity estimate. The vorticity of particle $a$
is thus

\begin{equation}
\omega_a = \left ( \nabla \times \mathbf{v} \right )_a = a_{2,1} - a_{1,2}.
\end{equation}

In order to calculate the linear velocity coefficients, a least squares method
was used to minimize the squared error between the actual SPH velocities of a
set of particles in the neighbourhood of particle $a$ (i.e. within $2h$), and
the linear estimate.

\begin{figure}[htbp]  \begin{center}
\subfigure[$t=0$]{\includegraphics[width=0.32\textwidth]{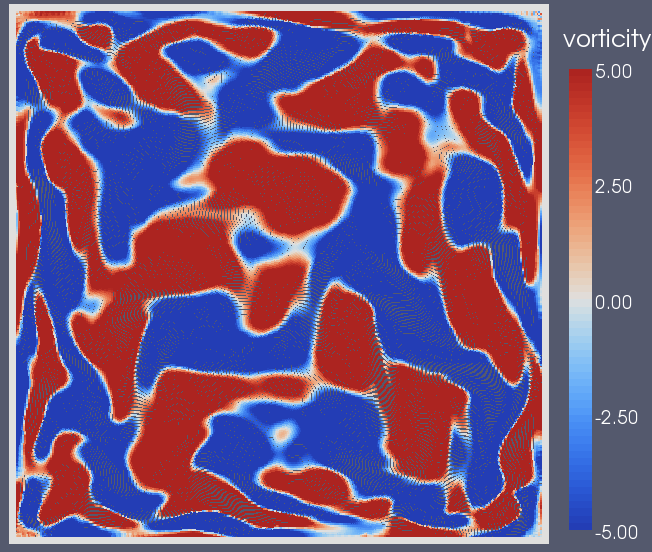}}
\subfigure[$t=3.2$]{\includegraphics[width=0.32\textwidth]{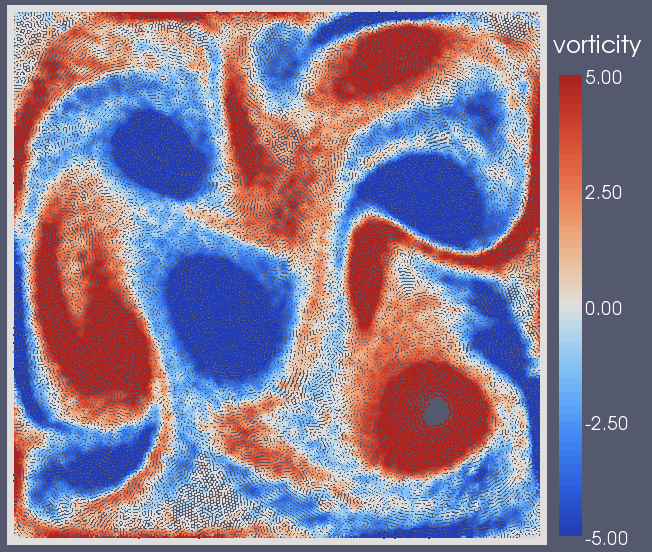}}
\subfigure[$t=6.3$]{\includegraphics[width=0.32\textwidth]{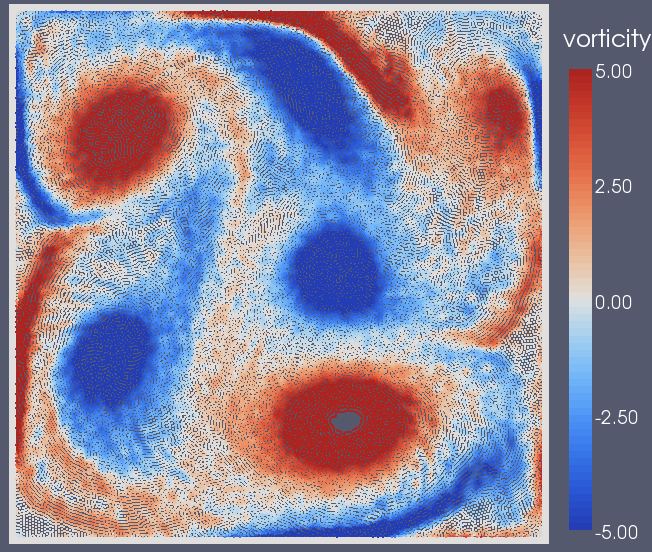}}
\subfigure[$t=12.6$]{\includegraphics[width=0.32\textwidth]{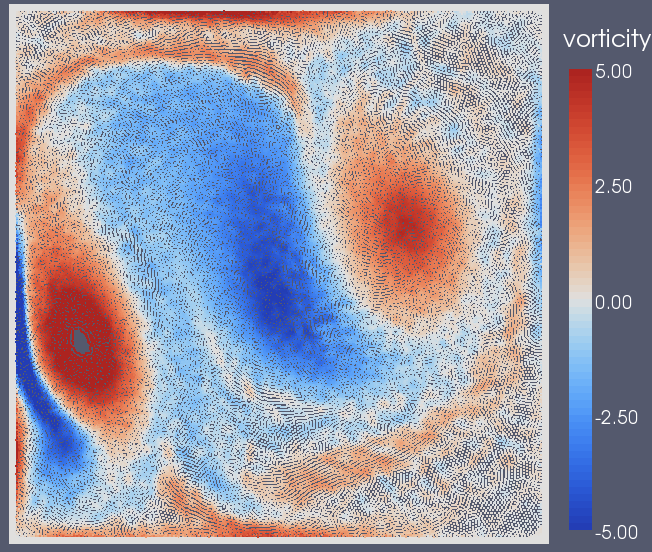}}
\subfigure[$t=25.2$]{\includegraphics[width=0.32\textwidth]{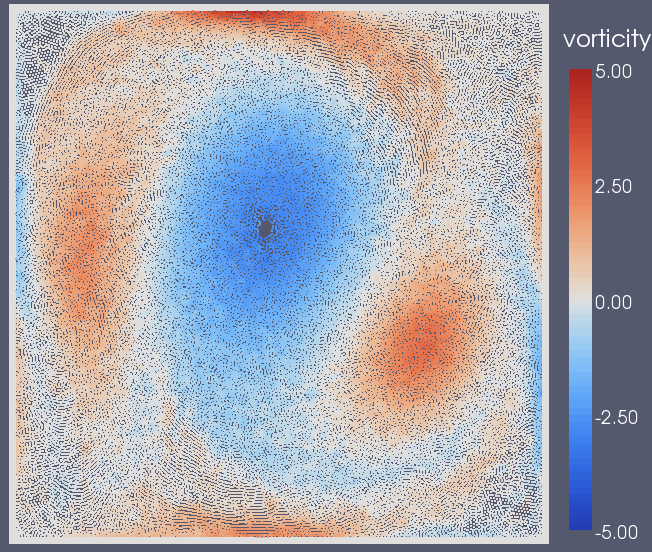}}
\subfigure[$t=37.8$]{\includegraphics[width=0.32\textwidth]{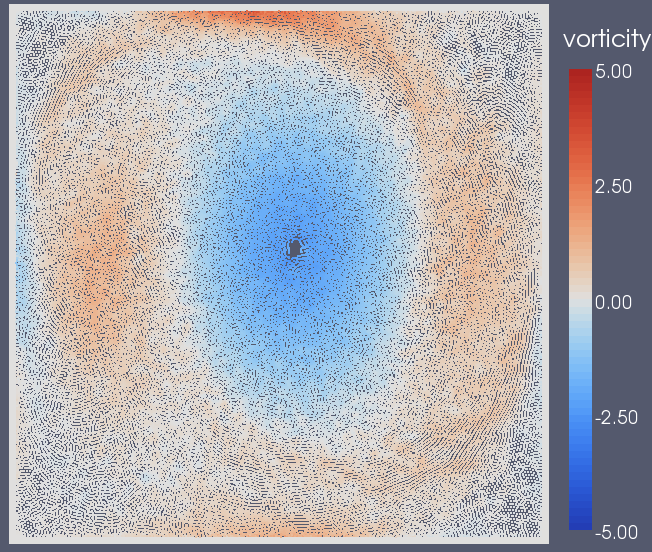}}
\caption{Evolution of a typical decaying turbulence simulation, which shows the
decay of the initial turbulent vorticity field to a single large vortex. The
particles are coloured by vorticity, which has been clipped to $\pm5$ in order
to show the field at later times.} \label{Fig:vorticityEvolution} \end{center} \end{figure}

Figure \ref{Fig:vorticityEvolution} shows a number of snapshots from the simulation
with the particles coloured by vorticity. The vorticity range has been clipped
between $\pm 5$ in order to show the vortex structure at later times. Figure
\ref{Fig:vorticityEvolution} shows the evolution of the vorticity field from
the initial turbulent field to the final state, a single large vortex taking up
most of the domain. Note that this is an example of the most likely evolution
of the turbulence field, corresponding to a strong spin-up (i.e. increase in
total angular momentum) of the fluid into a single large vortex.
Clercx et al. \cite{clercx99Decaying2DTurbinSquareCont} reported three main categories of
evolution, the other two being a weak or delayed spin-up and no spin-up at all.
This last case is characterised by having a final state consisting of a pair of
vortices with opposite spin. See Section
\ref{Sec:DecayTurbTotalAngularMomentum} for more details.

The initial evolution of the vorticity field occurs quite rapidly.  Vortices
usually either merge with neighbours with a similar rotation direction or are
stretched out into thin vortex filaments (and ultimately erased through the
action of viscosity) by vortices with an opposite spin. More rarely, a large
vortex can be split by a counter-rotating vortex into two smaller but still
coherent vortices. After $t\approx20$ the flow has settled down enough so that
there are only a few vortices left, and typically one monopole vortex that is
comparable to the size of the domain. This monopole vortex is also the
indicator of the spontaneous increase in total angular momentum for the flow,
also known as the spontaneous spin-up.  Subsequent to this, the other, smaller
vortices gradually fade away until only the single, large vortex remains, along
with the boundary layers that it excites along the surrounding walls. 

Most of these snapshots show strong boundary layers being generated by vortex
interactions with the walls. These are lifted away from the boundary to form
vortex filaments. However, these are generally short-lived and do not persist
as coherent vortices. Clercx et al. \cite{clercx99Decaying2DTurbinSquareCont} reported
that the vortex filaments contributed significantly to the turbulence over
$1<t<20$. The filaments are generated at the boundary and move into the flow to
become either elongated vorticity filaments or would be rolled up to form a
coherent vortex. This vortex would often pair up with an existing vortex to
form a dipole structure. In the SPH results only the first of these two options
is seen. The vorticity filaments are produced at the boundary and lifted away
to briefly form elongated filaments before they are erased by viscosity. Very
occasionally the filament would roll up to become a roughly round vortex,
however these structure were always much weaker than the surrounding structures
and were always quickly dissipated.

\section{Total Angular Momentum and Spontaneous Spin-Up}\label{Sec:DecayTurbTotalAngularMomentum}

\begin{figure}[htbp] 
\begin{center}
\includegraphics[height=0.8\textwidth,angle=-90]{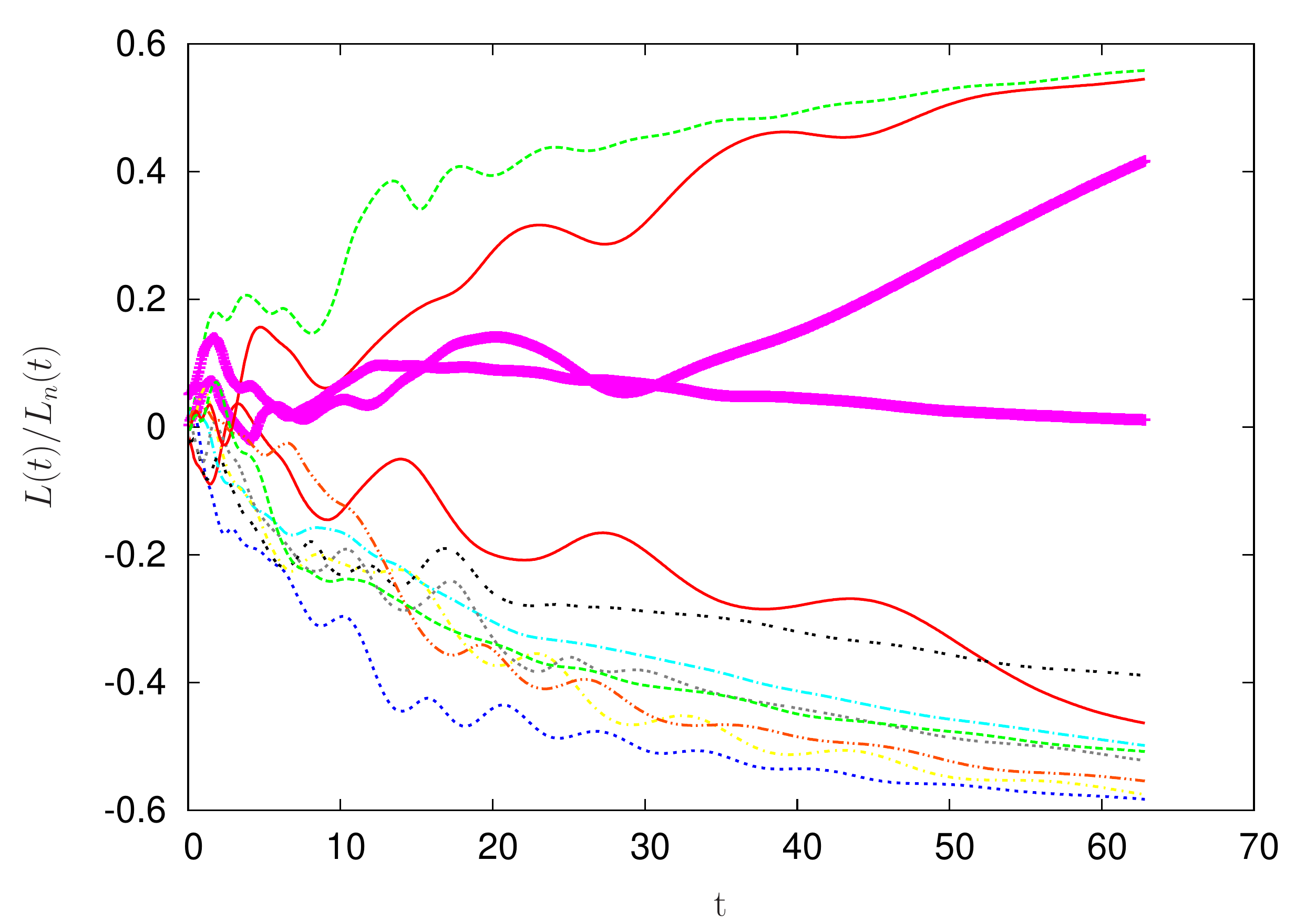} 

\caption{Each plot gives the total normalised angular momentum for each
ensemble run. The normalised angular momentum is given by $L(t)/L_n(t)$, where
$L_n(t)$ is the angular momentum of an equivalent mass of fluid with the same
kinetic energy $E(t)$ moving in rigid body rotation. That is, $L_n(t) = \sqrt{16
\rho \ell^4 E(t)/3}$} \label{Fig:totalAngularMomentum} 

\end{center} 
\end{figure}

Figure \ref{Fig:totalAngularMomentum} shows the normalised total angular
momentum $\tilde{L}(t)=L(t)/L_n(t)$ for 12 randomly initialised simulations.
The total angular momentum $L(t)$ is calculated around the centre of the box and is normalised by $L_n(t)$, the angular
momentum of an equivalent mass of fluid with identical kinetic energy $E(t)$
moving in rigid body rotation (i.e. $L_n(t)=\sqrt{16 \rho \ell^4 E(t)/3}$). Of the 12
simulations, 8 of them show a rapid, spontaneous spin-up that was clearly
established by $t \approx 20$. A further 2 simulations showed a slightly slower
spin-up (solid red lines), and the final 2 simulations show little or no
spin-up (thicker purple lines). This is identical to the ratio reported by
Clercx et al. in their pseudo-spectral results. Out of 12 simulations, 8 showed
a strong spin-up, 2 a weak spin-up and the final 2 showed no spin-up. 

This strong spin-up is the primary characteristic of turbulence in a square
wall-bounded domain. Due to the geometry of the boundary, the pressure forces
normal to the wall exert a net torque on the flow. The particular direction of
this torque is highly dependant on the initial velocity field. We have used an
initial field with an angular momentum close to zero and hence different random
fields produce a spin-up in either direction. Other decaying turbulence
experiments \cite{heijst06effects} have shown that simulations with a net
initial angular momentum can still spin-up in either direction, although if
the spin-up was in the opposite direction to the initial value then the
strength of the final spin-up is reduced accordingly. 

The normalised angular momentum plots given by Clercx et al.
\cite{clercx99Decaying2DTurbinSquareCont} indicate that the simulations with a
strong spin-up will converge to an angular momentum of $|\tilde{L}|=0.6$. This is
consistent with the SPH results. 

\begin{figure}[htbp]  \begin{center}
\subfigure[$t=0$]{\includegraphics[width=0.32\textwidth]{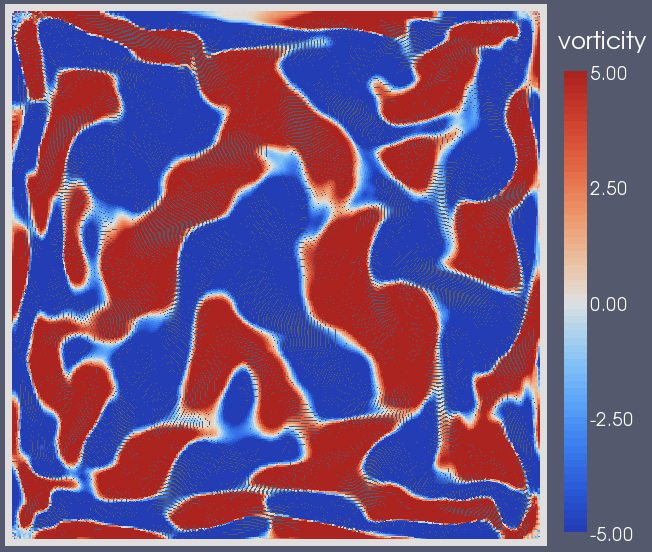}}
\subfigure[$t=3.2$]{\includegraphics[width=0.32\textwidth]{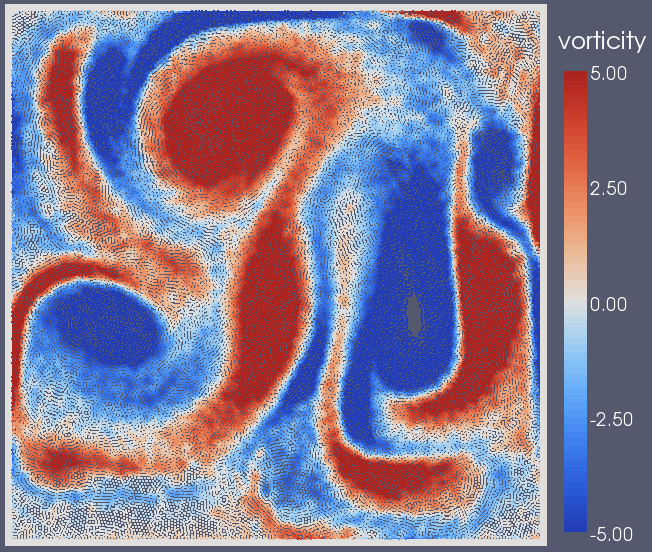}}
\subfigure[$t=6.3$]{\includegraphics[width=0.32\textwidth]{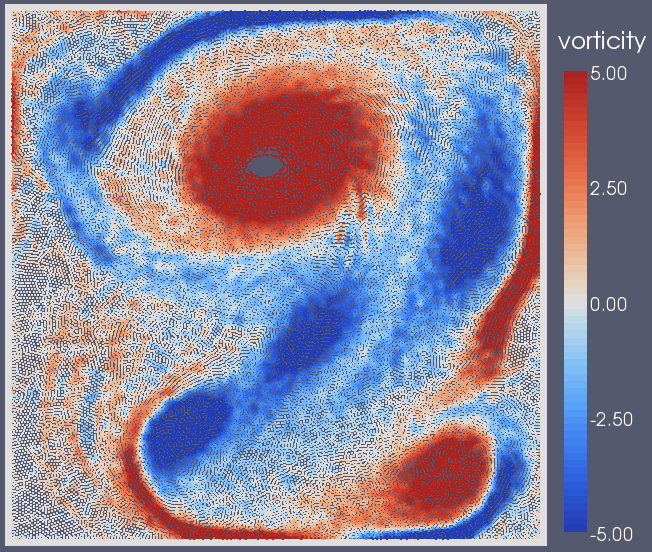}}
\subfigure[$t=12.6$]{\includegraphics[width=0.32\textwidth]{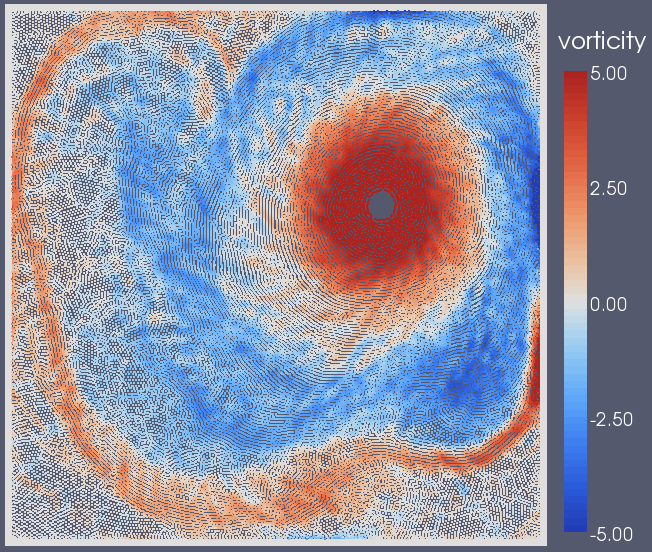}}
\subfigure[$t=25.2$]{\includegraphics[width=0.32\textwidth]{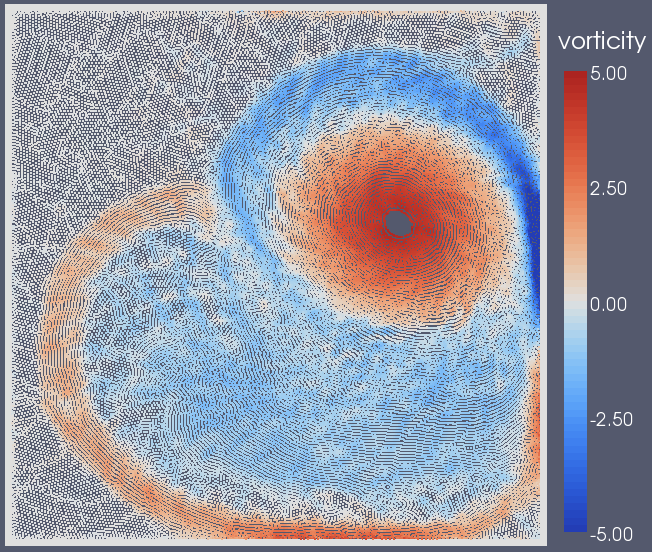}}
\subfigure[$t=37.8$]{\includegraphics[width=0.32\textwidth]{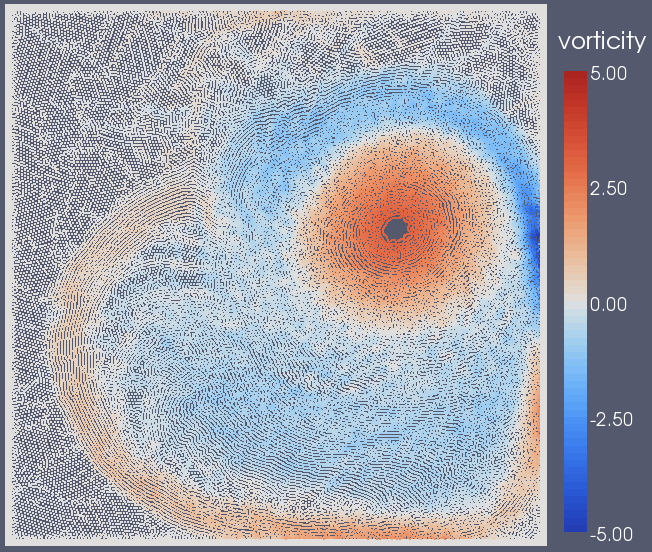}}
\caption{Evolution of a decaying turbulence simulation that shows the rarer
case where there is no spontaneous spin-up of the fluid.}\label{Fig:vorticityEvolutionNoSpinup} \end{center}
\end{figure}

Figure \ref{Fig:vorticityEvolutionNoSpinup} shows the vorticity evolution for
one of the SPH simulation that showed little or no spin-up. As is
characteristic of these cases, instead of evolving into a single monopole
vortex the initial turbulence instead decays to a dipole vortex structure. This
is seen most clearly in the snapshot at $t=25.2$. There is still obviously a
stronger vortex of the two, but the dipole structure persists long enough to
only cause a weak spin-up. Looking back at the angular momentum plots in Figure
\ref{Fig:totalAngularMomentum}, this vorticity evolution corresponds to the
thick purple $\tilde{L}(t)$ plot that shows no spin-up until $t=30$ when
$\tilde{L}(t)$ suddenly starts to increase. At this time the weaker vortex of
the dipole dissipates, allowing the remaining vortex to induce the spontaneous
spin-up.
\\
\\

\section{Kinetic Energy Decay}

\begin{figure}[htbp] 
\begin{center}
\includegraphics[height=0.8\textwidth,angle=-90]{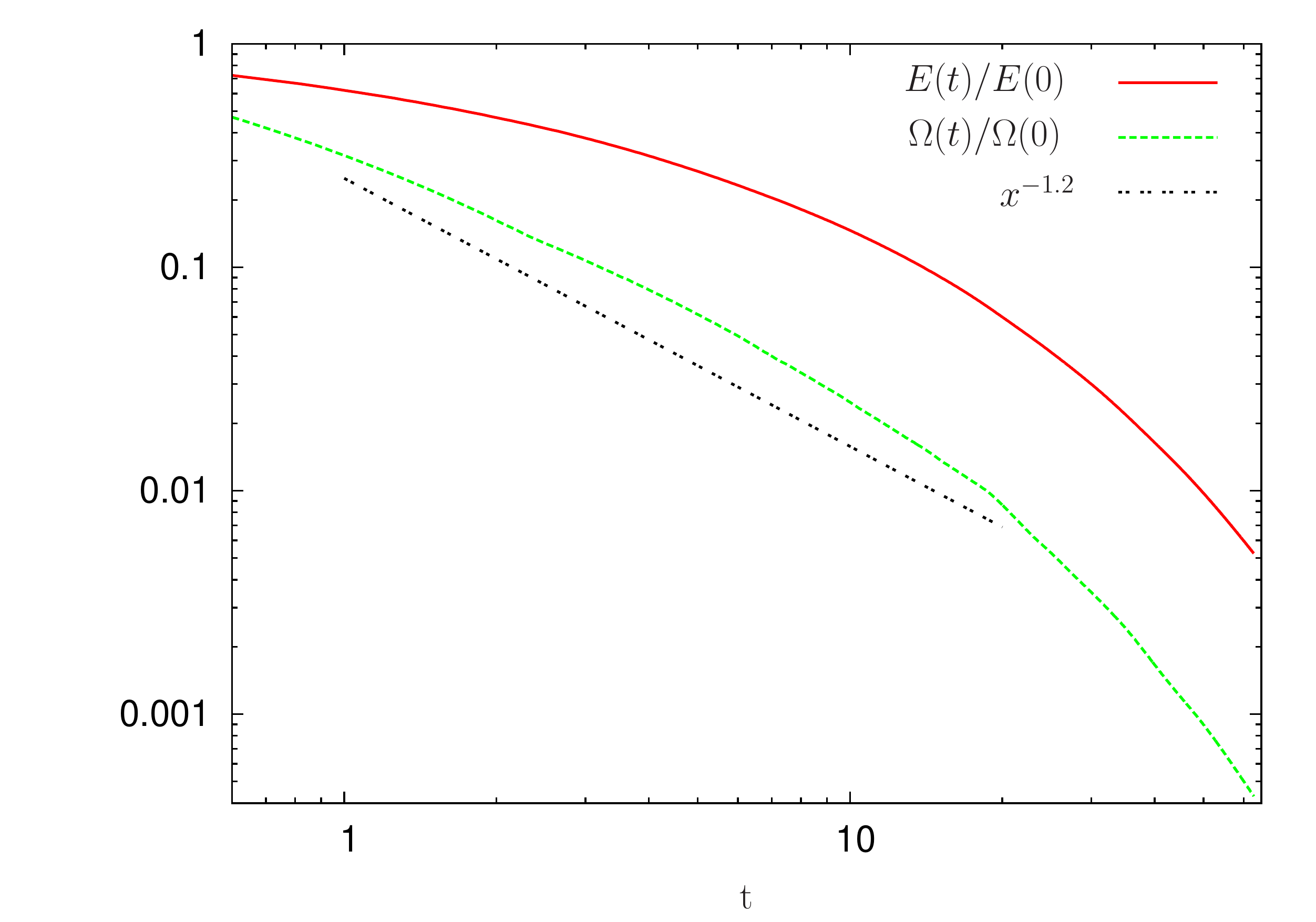}
\caption{Ensemble average of total normalised kinetic energy and enstrophy for all 12 decaying turbulence simulations.}\label{Fig:kineticEnergyDecay} 
\end{center} 
\end{figure}

Figure \ref{Fig:kineticEnergyDecay} shows the variation of total kinetic energy
and enstrophy with time. These ensemble statistics were calculated by averaging
over the 12 different simulation runs. They are both normalised by their values
at $t=0$. That is, $\tilde{E}(t)=E(t)/E(0)$ and
$\tilde{\Omega}(t)=\Omega(t)/\Omega(0)$.

The decay of kinetic energy and enstrophy is rapid over the time period from
$t=0$ to $t\approx20$. During this time the flow is turbulent and there are
many dissipative interactions between the vortices. Initially, the decay of
enstrophy is significantly greater than the kinetic energy. This is expected from a
two-dimensional turbulent flow, where the kinetic energy decay will approach zero as
$Re\rightarrow \infty$. As the simulated turbulence decays, the Reynolds number
of the flow decreases and the decay rate of the enstrophy and kinetic energy
become comparable. After $t\approx20$ the turbulence has evolved into a single
monopole vortex (in most cases), and the kinetic energy and enstrophy decay level out to a roughly
exponential decay.

The experimental results by Maassen et al. \cite{maassen02self} show a power
law decay for both kinetic energy and enstrophy for $1<t<20$.  The kinetic
energy decay shown in the SPH simulations does not follow this power law.
However, Maassen et al. state that 3D stratification effects in their
experiments had a large impact on the kinetic energy dissipation of the
turbulence. The kinetic energy decay shown here is closer in form to the
pseudo-spectral 2D turbulence simulations shown by Clercx et al.
\cite{clercx99Decaying2DTurbinSquareCont}. While Clercx et al. do not provide
any curve fitting for their kinetic energy decay plots, they are qualitatively
similar to the SPH results. In addition, both simulations have the same
average decay rate over $0<t<60$. For $Re=1500$, their pseudo-spectral results
showed a decay from $\tilde{E}=1$ to $\tilde{E}=0.001$ over approximately 60
time units. This is consistent with the SPH results.

The initial SPH enstrophy decay is similar to the power law seen in the
experimental results by Maassen et al. \cite{maassen02self}. Maassen et al. state
that, with a weak stratification, the enstrophy decay over $1<t<20$ is
proportional to $t^\alpha$, where $\alpha=-1.5\pm0.2$. The decay of the SPH
enstrophy over the same time period fits a power law with $\alpha=-1.2$. This
is slightly less than the experimental results, however, Maassen et al. also
show a slight decrease in $\alpha$ with decreasing stratification, indicating
that 3D effects in the experiments act to increase the enstrophy dissipation.
Therefore it can be concluded that the SPH enstrophy decay is consistent with
these experimental results.  

\section{Average Wavenumber Decay}

In Fourier or wavenumber space the total enstrophy $\Omega(t)$ is related to the kinetic energy spectrum by

\begin{equation}
\Omega(t) = \frac{\pi^2}{L^2 M}\int_0^\infty k^2 E(k) dk.
\end{equation}

where $E(k)$ is the kinetic energy at the integer wavenumber $k$ (defined such that the
wavelength of each mode is $\lambda = \frac{2L}{k}$) and $M$ is the total mass. It is reasonable to
define an "average" squared wavenumber $\langle k^2 \rangle$ as
\cite{clercx99Decaying2DTurbinSquareCont}

\begin{equation}
\langle k^2 \rangle = \frac{\Omega(t)}{E(t)} \sim \frac{\int_0^\infty k^2 E(k) dk}{\int_0^\infty E(k) dk}
\end{equation}

\begin{figure}[htbp]  
\begin{center}
\includegraphics[height=0.8\textwidth,angle=-90]{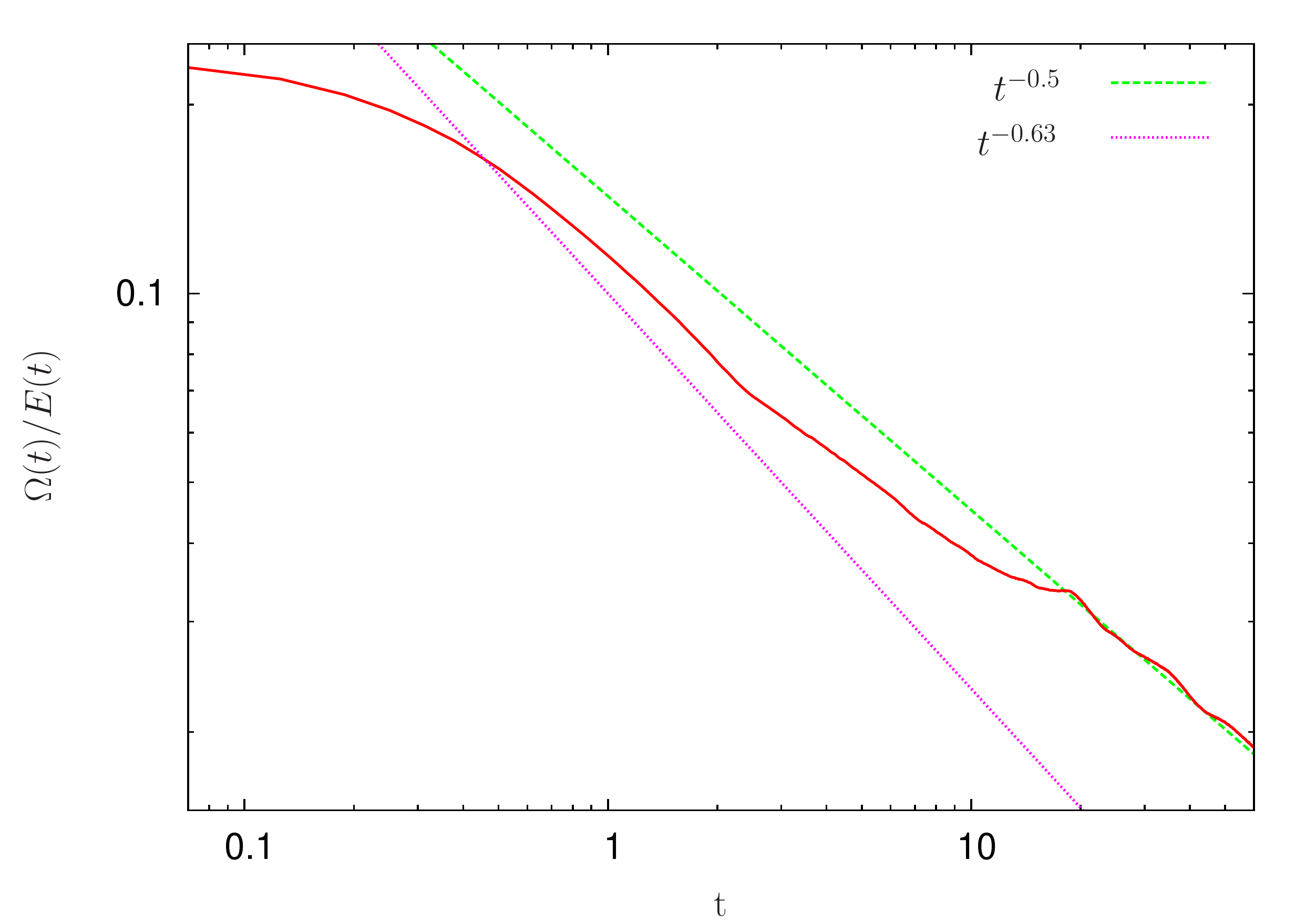} 

\caption{Ensemble average of $\Omega(t)/E(t)$, where $E(t)$ is the total
kinetic energy and $\Omega(t)$ is the total enstrophy. This gives an estimate
of the mean squared wavenumber $\langle k^2 \rangle$. The higher
dashed green reference line shows the average slope of the SPH results for
$t>20$, the lower dashed violet reference line shows the scaling reported by
Clercx et al. \cite{clercx99Decaying2DTurbinSquareCont} over the time period $0.7<t<10$.}
\label{Fig:averageWavenumberDecay}
\end{center} 
\end{figure}

This variable is an indication of the average squared size of the eddies
present in the turbulence. Taking an ensemble average of the average wavenumber
over all 12 SPH simulations gives the plot shown in Figure
\ref{Fig:averageWavenumberDecay}. Also shown in this plot are two reference
lines. Clercx et al. \cite{clercx99Decaying2DTurbinSquareCont} found that the decay in
$\langle k^2 \rangle$ was approximately $t^{-0.63}$ from $t\approx0.7$ to
$t\approx10$. This scaling is depicted by the lower, purple reference line. The
decay in $\langle k^2 \rangle$ for the SPH results is slightly less than this
over the same time period, indicating that the evolution of the kinetic energy
into larger vortex structures is slightly slower for the SPH simulations. In
other words, the strength of the inverse energy cascade is weaker. 

After $t = 20$ the turbulence clearly enters a different evolutionary stage.
Leading up to this time, over $10<t<20$, the decay in the average wavenumber
begins to level off. At $t= 20$ this trend is suddenly reversed and $\langle
k^2 \rangle$ resumes decaying. After this time a significant amount of
fluctuation is introduced to the plot of $\langle k^2 \rangle$. The rate of
decay for $t>20$ is approximately $t^{-0.5}$, as shown by the higher, green
reference line. Both the behaviour of the average wavenumber at $t=20$ and a similar
rate of decay for $t>20$ is also seen in the pseudo-spectral results by Clercx
et al.

This sudden change in $\langle k^2 \rangle$ signals the formation of the large
monopole vortex that is the typical end-state of these simulations. The
interesting point to make about the dominant "kink" in $\langle k^2 \rangle$ is
that it is not smoothed out due to the ensemble averaging over the 12
simulations. All these simulations have different initial velocity fields that
lead to quite different evolutions for the turbulent flow.  As we have already
noted, the coherent vortices and hence the germination of the large monopole
vortex is always embedded in this initial velocity field.  It is therefore
interesting that this vortex (or the dipole vortex in the case of no spin-up)
is formed at the same time in each simulation, and its effects on the average
wavenumber of the flow are so consistent across the ensemble simulations. 

\section{Kinetic Energy Spectrum}\label{Sec:KineticEnergySpectrum}

In the late sixties Batchelor \cite{batchelor69energySpectrum} and Kraichnan
\cite{kraichnan67inertialRange} published their seminal papers on
two-dimensional turbulence. They proposed the idea of an enstrophy cascade, in
which the interactions between eddies resulted in a movement of enstrophy to
increasingly higher wavenumbers. Using the assumption that the interactions of
the turbulent eddies were local in wavenumber and that the kinetic energy
spectrum is self-similar over time and space, they derived the following
scaling law for the kinetic energy spectrum.

\begin{equation}\label{Eq:turbDirectEnstrophyScaling} 
E(k) \sim \beta^{2/3}k^{-3}.
\end{equation}

Due to the walls of the box, Batchelor and Kraichnan's assumption of isotropic
and homogeneous turbulence is no longer valid and therefore the classical
$k^{-3}$ enstrophy cascade spectrum is not necessarily expected. However,
numerical experiments of decaying turbulence (from an initial regular array of
gaussian vortices) in a no-slip box \cite{clercx00energy} have provided some
data for comparison. In this paper, Clercx and van Heijst compare the kinetic
energy spectrum of periodic and wall-bounded turbulence, using a one
dimensional spectrum obtained by a Chebyshev expansion of the kinetic energy
and averaging over specific $x$ and $y$ coordinates to distinguish between the
centre of the box and the neighbourhood of the no-slip walls. Near the no-slip
boundaries, the authors show that there is an absence of a direct enstrophy
cascade and that the kinetic energy spectrum is much flatter in the inertial
range than the periodic case (the spectrum scales as $k^{-5/3}$ for $Re=20,000$
and $k^{-2.4}$ for $Re=5000$). Further away from the boundaries, they show that
a $k^{-3}$ inertial range quickly forms near the centre of the box. However, in
contrast to the periodic case, this spectrum flattens slightly over time,
evolving to $k^{-2.1}$ for $Re=20,000$ or $k^{-2.5}$ for $Re=5000$.

In order to compare against the SPH results, a method is needed to perform a Fourier transform
over the SPH particles. Since they will be unevenly distributed, traditional
Fast Fourier Transform methods are of no use unless the particle variables are
first interpolated to a grid, which would artificially reduce the spectrum at
large wavenumbers. 

Mansour \cite{mansour07SPH} proposed a SPH Fourier transform by taking the usual form of the Fourier
transform and simply replacing the integral with an SPH sum over all the
particles. For example, a 2D Fourier transform over a periodic square of side $2L$ is

\begin{align}
F(\mathbf{k}) &= \frac{1}{L^2} \int_{-L}^{L} \int_{-L}^{L} f(\mathbf{r})e^{-i\frac{\pi}{L} \mathbf{k} \cdot \mathbf{r}}dx dy,
\end{align}

where $\mathbf{k}$ is an integer wavevector and $f(\mathbf{r})$ is the function
being transformed. Converting this equation into an SPH summations, this
becomes

\begin{align}\label{Eq:SPHfourierTransform}
F(\mathbf{k}) &= \frac{1}{L^2} \sum_{b=1}^{N} f_b e^{-i\frac{\pi}{L} \mathbf{k} \cdot \mathbf{r}_b}\frac{m_b}{\rho_b},
\end{align}

where $\mathbf{r}_b$, $m_b$ and $\rho_b$ are the position, mass and density of
particle $b$ and $f_b$ is the variable that is being transformed to the Fourier
domain. The 2D spectrum $F(\mathbf{k})$ is then collapsed to a 1D spectrum
by averaging over the angles in wavenumber space. That is

\begin{equation}
F(k) = \frac{1}{\sum_{k-1 < |\mathbf{k}| \le k} 1} \sum_{k-1 < |\mathbf{k}| \le k} |F(\mathbf{k})|,
\end{equation}

where $k = |\mathbf{k}|$.

Mansour \cite{mansour07SPH} tested the accuracy of this SPH Fourier transform in 2D
using a number of test spectra and density profiles and found that accurate
results were obtained up to wavenumber $k_c=0.39N$, where $N$ is the number of
particles along that dimension. Additional tests by the authors
have shown that the transform is accurate up to wavenumber $k_c=0.26N$.
Consequentially, all the spectra shown in this paper are shown over the range
$0<k<0.26N$.

Due to the no-slip boundaries, the results will no longer satisfy
the periodic assumption. Following the example of Wells et al. \cite{wells07vortices}, the
variable to be transformed is first reduced by its mean and then a Hann window
is applied. Therefore, to calculate the kinetic energy spectra, the variable $f_b$ in (\ref{Eq:SPHfourierTransform}) is calculated using

\begin{equation}\label{Eq:HannWindow}
f_b = 
\begin{cases}
0.5(1-cos(\pi (|\mathbf{r}_b|+1)))(\frac{1}{2} m_b \mathbf{v}_b \cdot \mathbf{v}_b - \frac{\sum_a{\frac{1}{2} m_a \mathbf{v}_a \cdot \mathbf{v}_a}}{N}), & \mbox {for } |\mathbf{r}_b| \le 1, \\
0, & \mbox{ otherwise}.
\end{cases}
\end{equation}

\begin{figure}[htbp]  \begin{center}
\includegraphics[width=0.59\textwidth]{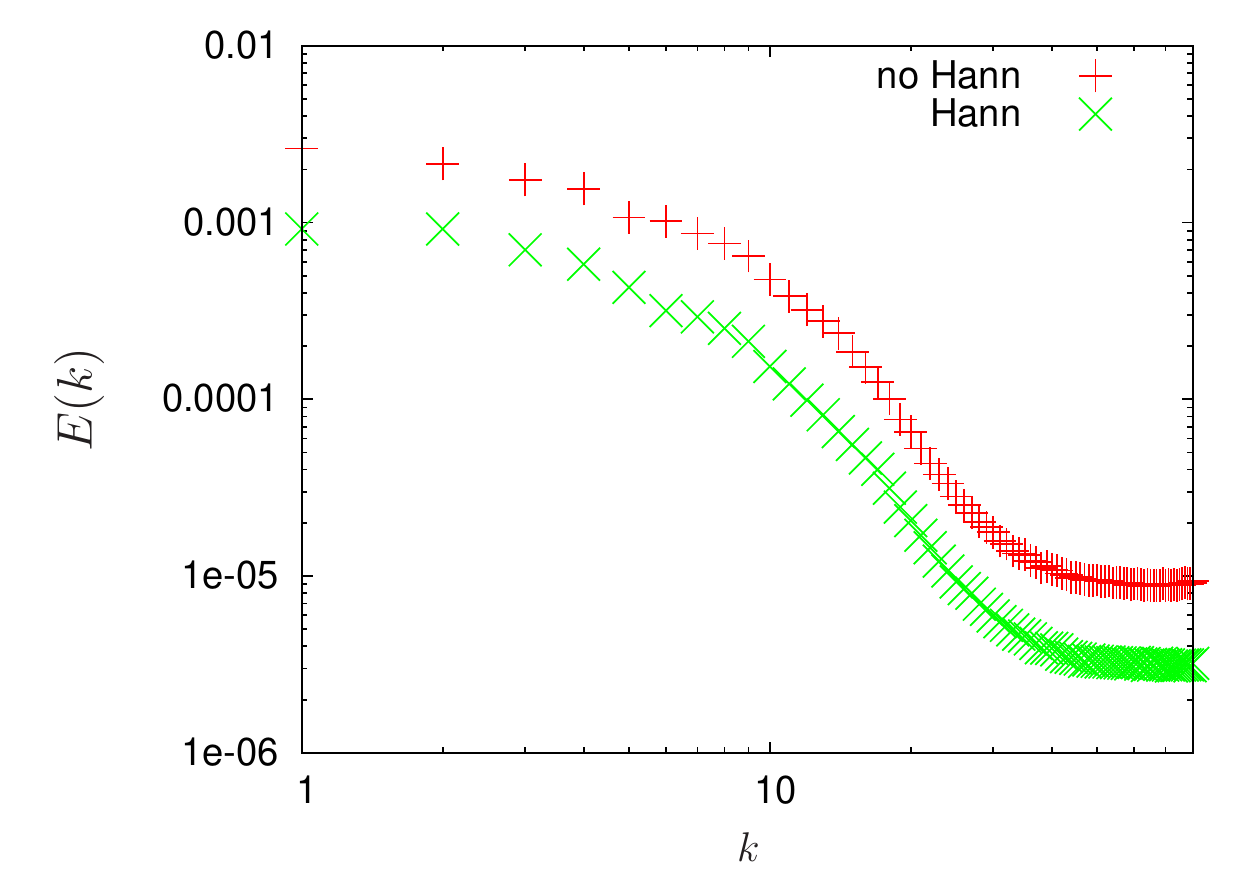}
\caption{Kinetic energy spectrum calculated using the SPH Fourier transform of
a periodic turbulent velocity field. This plot shows the effect of
pre-multiplying the kinetic energy by a Hann window (Equation
\ref{Eq:HannWindow}).} \label{Fig:hannNoHann} \end{center} \end{figure}

In order to determine the effect of the Hann window on the spectrum, a
turbulent velocity field was taken from a previous DNS SPH simulation of forced
turbulence in a periodic domain. In this case, the assumption of periodicity is
valid and the kinetic energy spectrum can be calculated without the Hann
window. The result is shown in Figure \ref{Fig:hannNoHann}, along with the spectrum
calculated after pre-multiplying by the Hann window. 

The original spectrum is shown using red plus
symbols. The spectrum calculated by pre-multiplying by the Hann window is shown
using green crosses, and is approximately one third of the original spectrum
over all the wavenumbers calculated. Therefore, it can be concluded that the
windowing does not significantly bias the shape of the spectrum, nor will it
have any effect on the $k^{-3}$ enstrophy cascade scaling law.

\begin{figure}[htbp]  \begin{center}
\subfigure[$t=0.63$]{\includegraphics[width=0.49\textwidth]{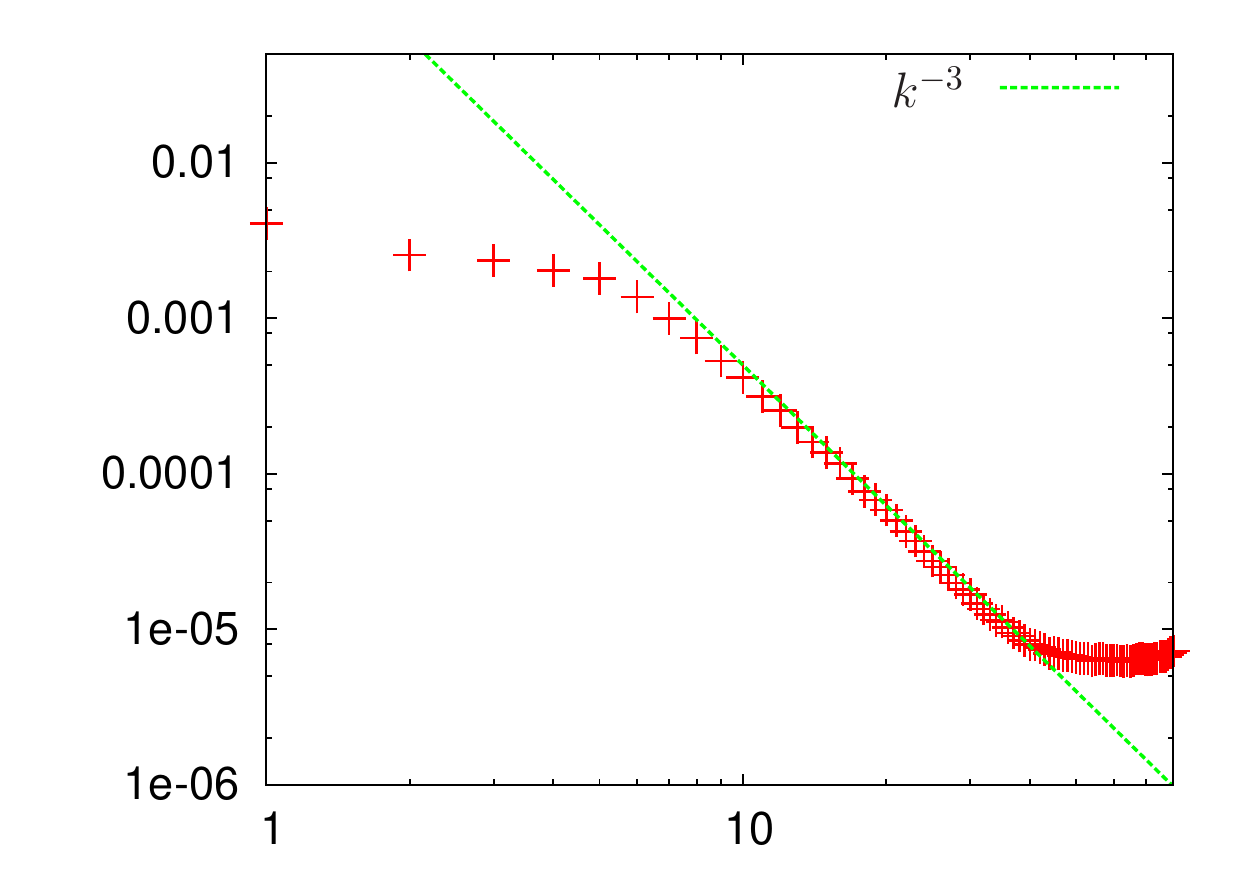}}
\subfigure[$t=3.77$]{\includegraphics[width=0.49\textwidth]{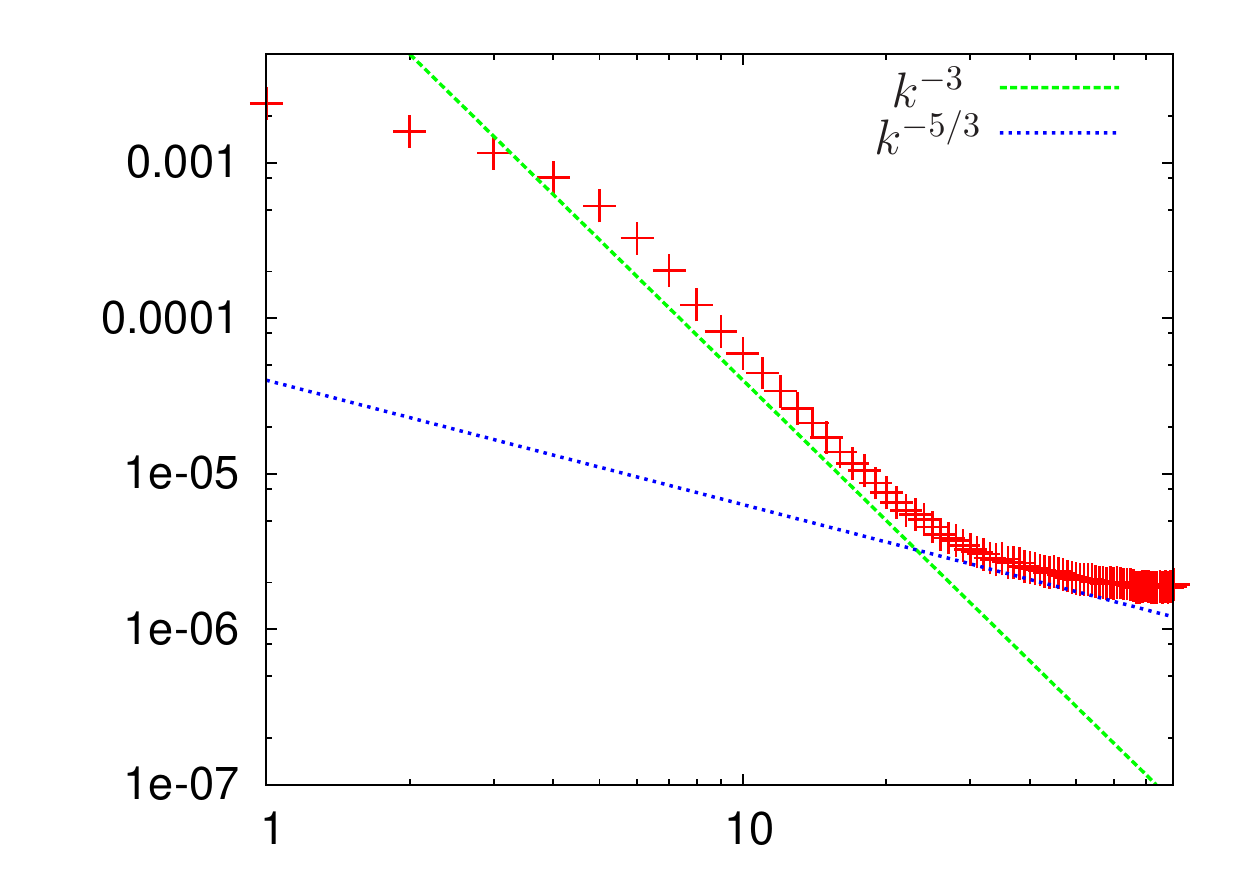}}
\subfigure[$t=9.42$]{\includegraphics[width=0.49\textwidth]{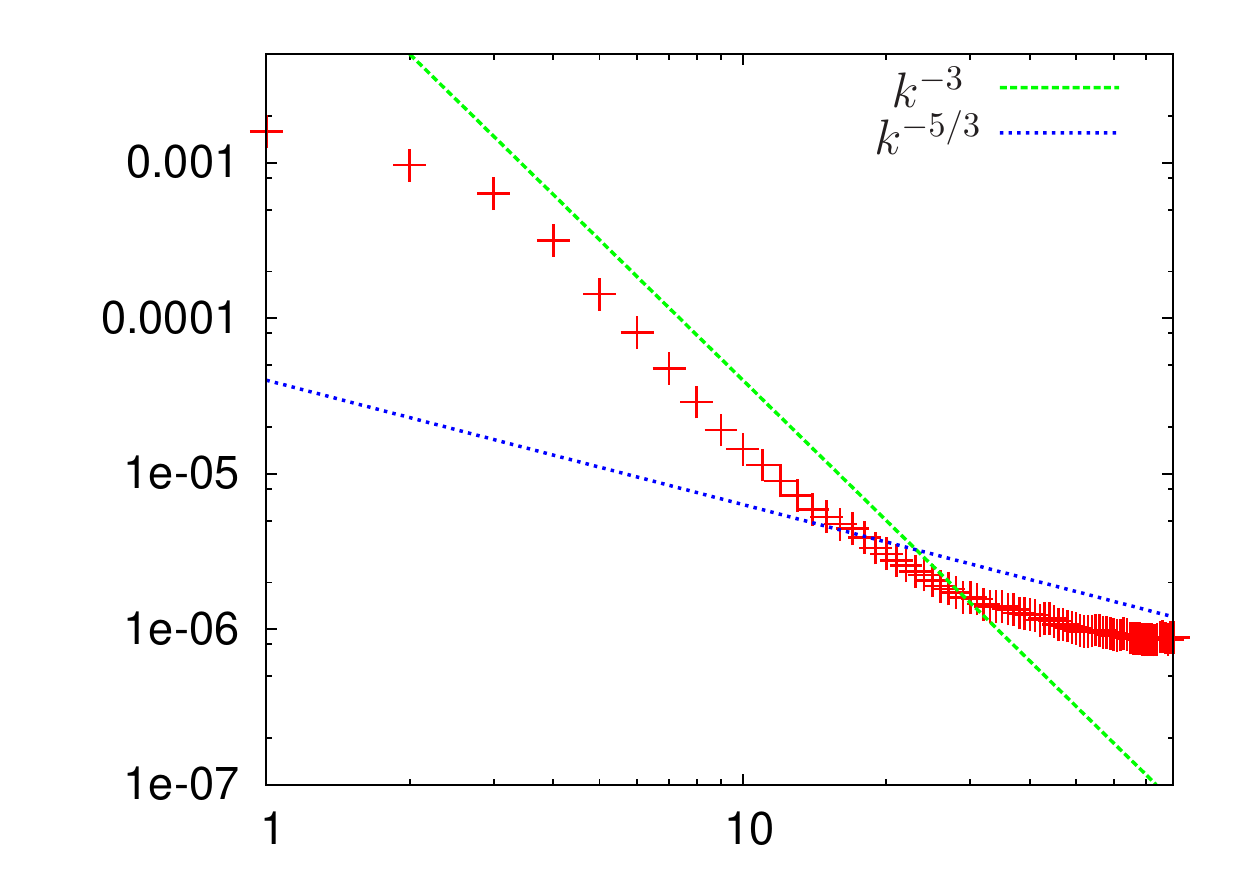}}
\subfigure[$t=21.99$]{\includegraphics[width=0.49\textwidth]{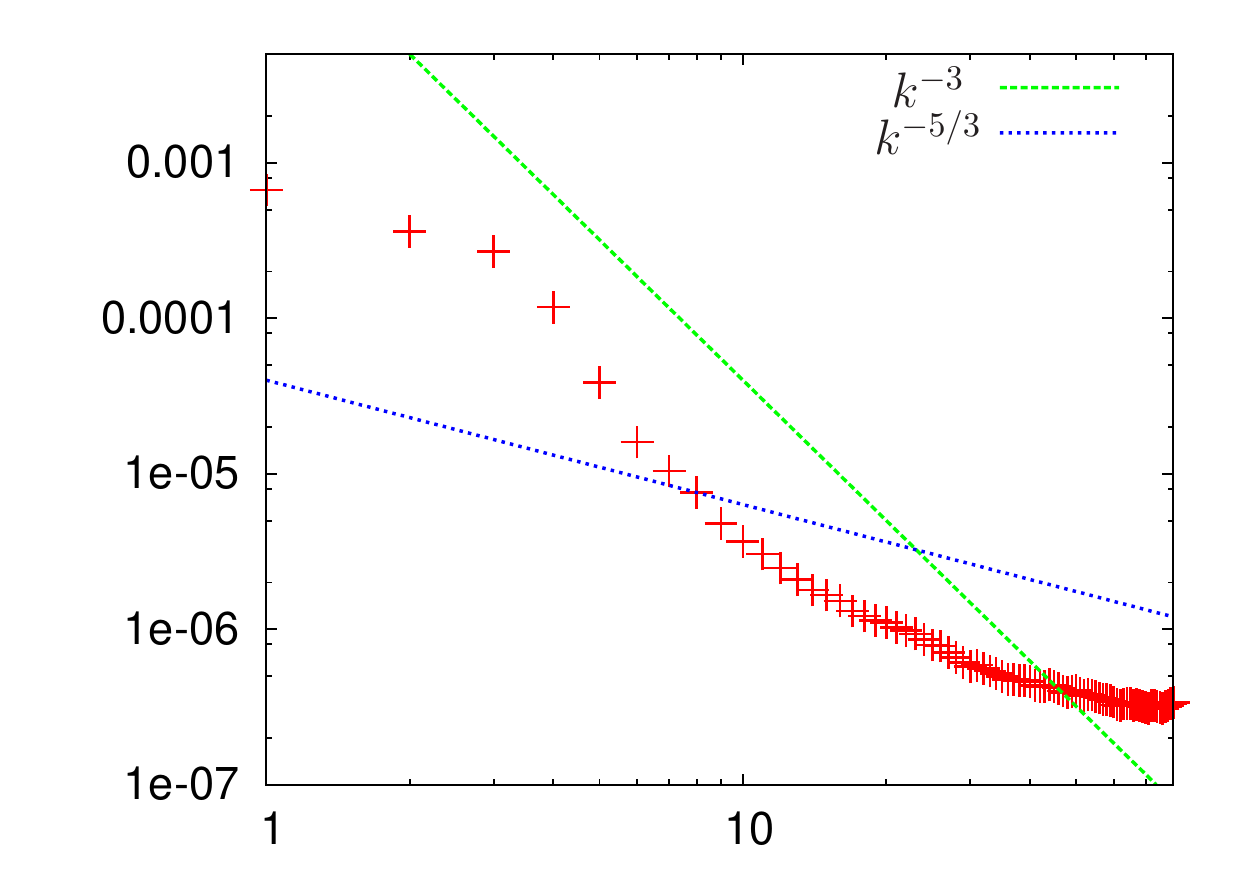}}

\caption{Evolution of the kinetic energy spectrum $E(k)$ during the formation
of the single monopole or dipole vortex. Each spectrum is plotted against the wavenumber
$k$.}\label{Fig:kineticEnergySpectrum} \end{center}

\end{figure}

Using this method for finding the SPH Fourier transform, the kinetic energy
spectrum of the SPH results have been calculated at four different timesteps
during the evolution of the decaying turbulence over $0<t<22$. This period
covers the decay of the turbulence to the final domain-filling monopole or
dipole vortex. Figure \ref{Fig:kineticEnergySpectrum} shows the resultant
spectra, which have been averaged over the 12 ensemble simulations.

By $t=0.63$ the SPH simulation has established a clear $k^{-3}$ inertial range,
which matches the published pseudo-spectral results by Clercx and van Heijst
\cite{clercx00energy}.  The $k^{-3}$ range occurs at wavenumbers $7<k<40$. The
lower bound at $k=7$ is reasonable, and is an indication of the largest scale
of the initial velocity field. The upper bound of $k=40$ is unexpected, as the
inertial range (and the $k^{-3}$ scaling) should continue down to the
dissipation length scale. This flattening of the spectrum is not seen in the
pseudo-spectral simulations in \cite{clercx00energy}. Clercx and van Heijst do
report that the spectrum flattens slightly over time at higher wavenumbers, but
this process does not start before 2 vortex turn-over times (i.e. $t \approx 2$). 

For $t>0.63$, the flattening of the spectrum spreads to smaller wavenumbers as
the turbulence decays. The reference lines shown in plots (b)-(c) show a range
of approximately $k^{-0.8}$ that grows to $k>20$ by $t=21.99$. While a similar
flattening was reported in \cite{clercx00energy} for $t>4$, this effect was much 
more subtle. Instead of the $k^{-0.8}$ scaling seen in the SPH results
(for $Re=1500$), Clercx and van Heijst \cite{clercx00energy} report that the
spectrum flattens to $k^{-2.5}$ for $Re=5000$, and $k^{-2.1}$ for $Re=20,000$.

The results from the SPH kinetic energy spectrum are mixed. While the SPH
results show a clear direct enstrophy cascade for much of the inertial range,
as indicated by the $k^{-3}$ spectrum, the kinetic energy spectrum at high
wavenumbers does not match the pseudo-spectral simulations of
\cite{clercx00energy}. While this might be partially attributed to the differences in
the simulations, in terms of initial velocity field and Reynolds number, of
most concern is the deviation of the $k^{-3}$ spectrum shown at $t=0.63$. This
is seen at very small times, when the majority of the fluid in the box has not
interacted with the boundaries. At these times the kinetic energy
spectrum should match the classical direct enstrophy cascade $k^{-3}$ that is
predicted by 2D turbulence theory and normally seen in periodic 2D turbulence
simulations.

However, it must be noted that the evolution of the integral variables of the
SPH turbulence (e.g. total kinetic energy, enstrophy and angular momentum), do
not seem to be significantly affected by the excess kinetic energy at high
wavenumbers and correspond closely to the published pseudo-spectral results.
This is an indication that this excess energy is largely cosmetic in nature and
does not significantly affect the turbulence evolution, at least for the case
of decaying turbulence.

\section{Convergence Study}\label{Sec:ConvergenceStudy}

The numerical method SPH is based on both the integral interpolant (Equation
\ref{Eq:integralInterpolant}) and the summation interpolant (Equation
\ref{Eq:summationInterpolant}). The integral interpolant has a second order
accuracy with the smoothing length $\mathcal{O}(h^2)$. The accuracy of the
summation interpolant is more complicated and depends on the distribution of
the particles and hence the dynamics of the flow.

When performing a convergence study using SPH, it is important to vary both the
number of particles used in the simulation, as well as the ratio of smoothing
length to particle spacing $h/\Delta p$. Increasing the number of particles
while keeping $h/\Delta p$ constant decreases the smoothing length $h$, which
increases the accuracy of the integral interpolant while keeping the error due
to the summation interpolant constant. Of course, if it is this latter error
that is dominating the solution, increasing the number of particles while
keeping $h/\Delta p$ fixed will have little effect on the solution. 

\begin{figure}[htbp]  \begin{center}
\includegraphics[angle=-90,width=0.8\textwidth]{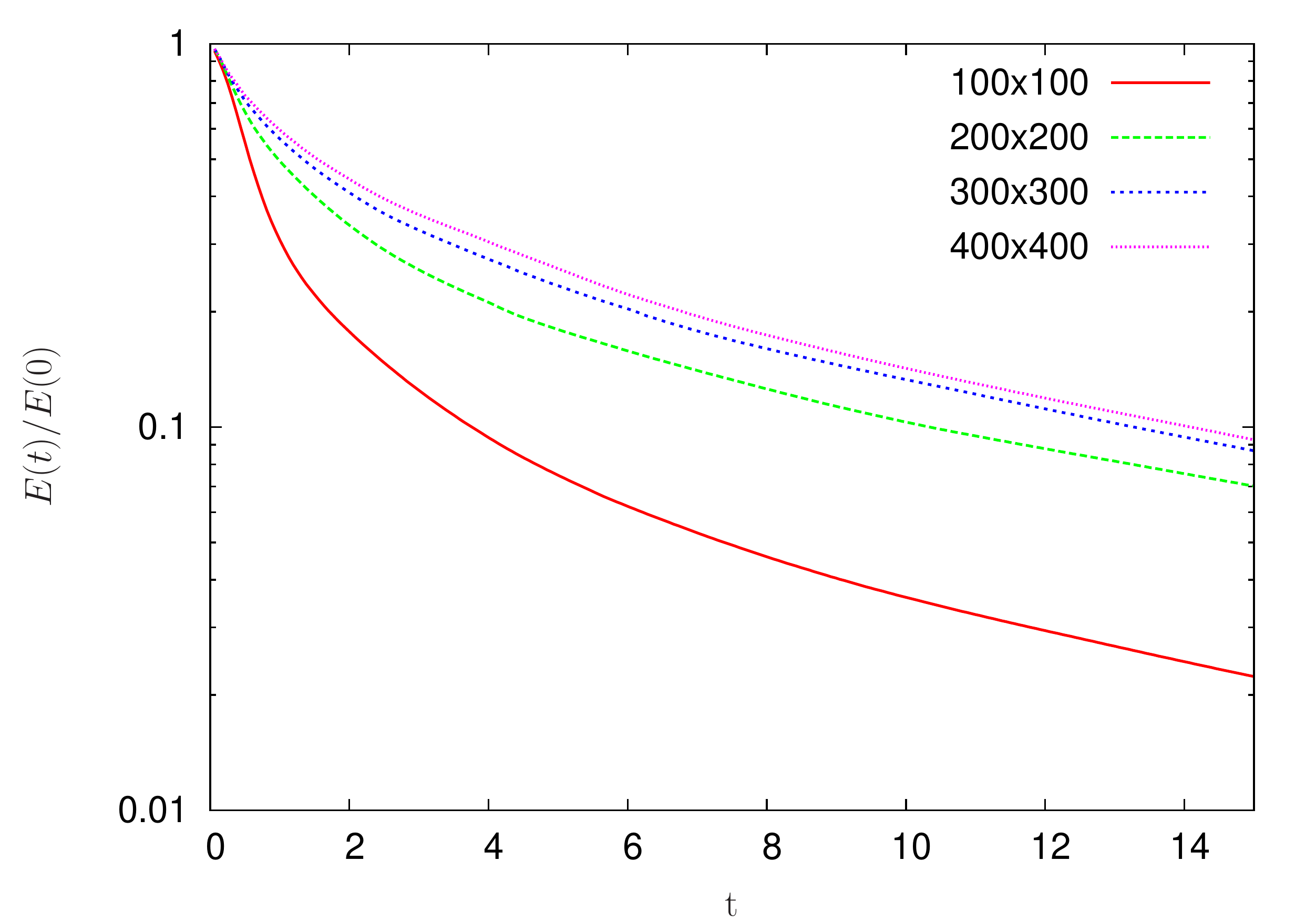}
\caption{Decay of total normalised kinetic energy $\tilde{E}(t)$ from the
same initial velocity field for different resolutions. $h/\Delta p$ is kept constant.}\label{Fig:decayKEconvergeRes} \end{center}
\end{figure}

Figure \ref{Fig:decayKEconvergeRes} shows the kinetic energy decay for a few
different resolutions. The ratio of smoothing length to particle spacing
$h/\Delta p$ is kept constant at 1.95 (all other SPH parameters, including the
Reynolds number, are also kept constant). At the lowest resolution the accuracy
of the simulation is poor and there is a substantial amount of numerical
dissipation. As the resolution is increased, this dissipation is reduced and
once the number of particles has increased beyond 300x300 there is no longer a
significant change in the results, indicating that the solution (or at least
this particular variable of the solution) has converged.

\begin{figure}[htbp] \begin{center}
\includegraphics[angle=-90,width=0.8\textwidth]{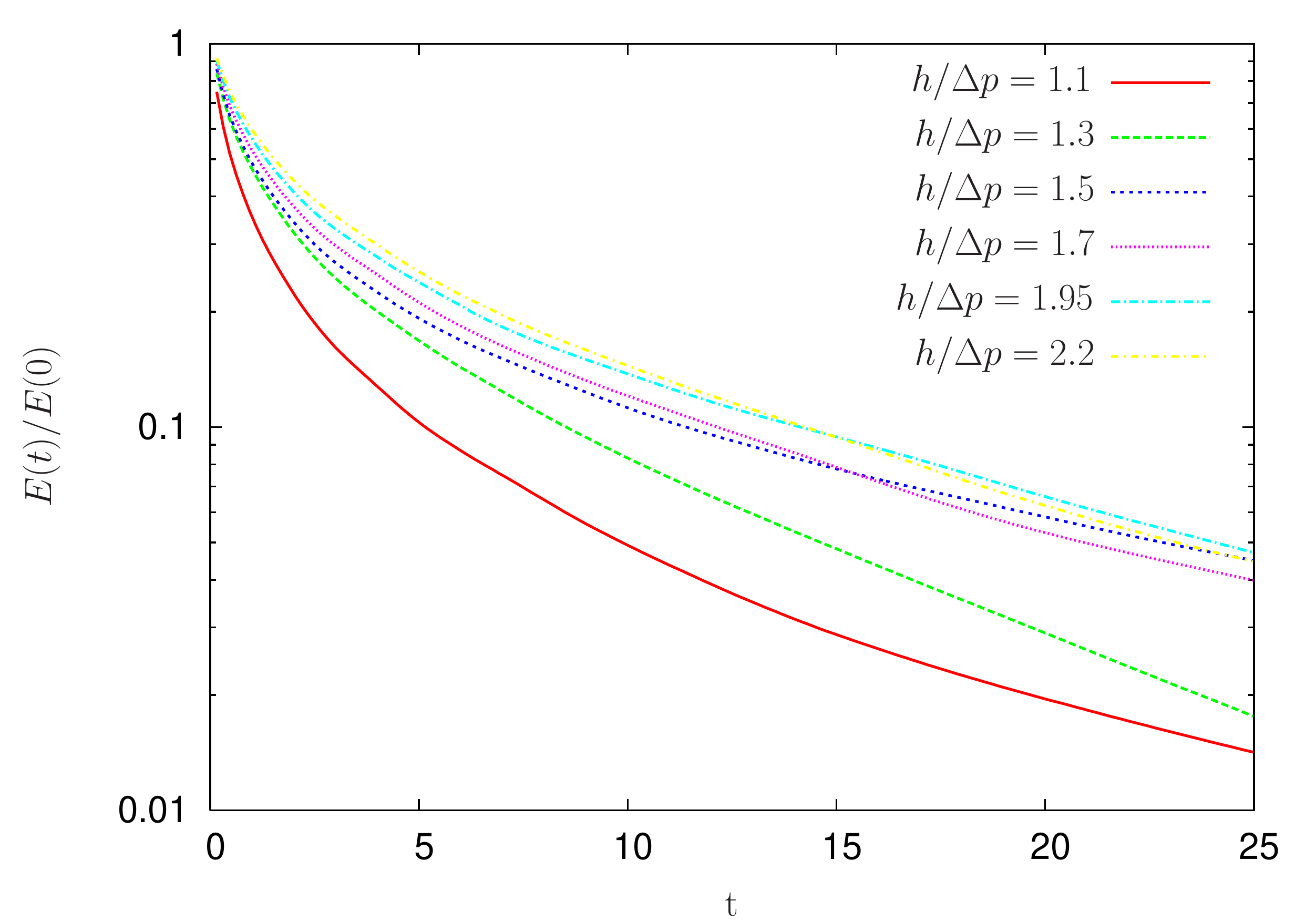}
\caption{Decay of total normalised kinetic energy $\tilde{E}(t)$ from the same
initial velocity field for different values of $h/\Delta p$. The smoothing
length $h$ is kept constant.}\label{Fig:decayKEconvergeHFAC} 
\end{center} 
\end{figure}

Figure \ref{Fig:decayKEconvergeHFAC} shows the same kinetic energy decay for
increasing values of $h/\Delta p$. In each simulation the number of particles
is also altered in order to keep $h$ constant at 0.013. This keeps the accuracy of the
integral interpolant constant while varying the error for the summation
interpolant. Using a value of $h/\Delta p \le 1.3$, the long term decay of kinetic energy is significantly enhanced due to
the excess dissipation caused by a noisy velocity field. However, setting
$h/\Delta p \ge 1.4$  dramatically reduced this error and brought the results
more into line with the pseudo-spectral kinetic energy decay rate. 

For $h/\Delta p \ge 1.4$, the long term evolution of the total kinetic energy
does not converge to a single decay rate. This is due to the chaotic nature of
the decaying turbulence. While all the simulations have an identical initial
velocity field, the variations in particle number and $h/\Delta p$ cause each
simulation to evolve into a different flow. In a similar manner to the angular
momentum results shown in Section \ref{Sec:DecayTurbTotalAngularMomentum}, some
simulations experience a strong spin-up while others only have a weak spin-up.

\section{Conclusion}
\vspace{2 mm}

The SPH results for decaying turbulence show a good agreement with all of the
quantitative pseudo-spectral results from Clercx et al.
\cite{clercx99Decaying2DTurbinSquareCont}. The decay of kinetic energy and the
characteristics of the spontaneous spin-up (i.e. it's strength and likelihood)
match well. The decay of the average squared wavenumber $\langle k^2 \rangle$
is consistent with the pseudo-spectral results, although the decay rate over
$0.7<t<20$ is slightly slower than that reported by Clercx et al. ($t^{-5}$ versus
$t^{-0.63}$). However, the SPH results do reproduce the clear phase change in
$\langle k^2 \rangle$ at $t\approx 20$ which signals the formation of the final
state for the turbulent decay. 

The most significant differences between the two simulations are qualitative in
nature. Clercx et al. \cite{clercx99Decaying2DTurbinSquareCont} emphasised the importance
that the boundaries play in injecting high intensity vorticity gradients into
the flow.  Clercx et al. showed that these vorticity filaments can roll up and
persist for long times, travelling far into the interior of the flow as
coherent vortices. This has been confirmed in both numerical experiment and
experimental results. Wells et al. \cite{wells07vortices} have even set up a forced
turbulence experiment where the flow was purely driven by the generation of
vorticity filaments at the boundaries. In the SPH simulations strong boundary
layers were generated at the boundaries in a similar manner to that described
by Clercx et al. However, once these were lifted away from the wall they
consistently failed to become long-lived structures. Instead, they were
elongated into filaments that would either be destroyed by viscosity or merge
with an interior vortex with an identical sign. None of the filaments rolled up
to become coherent vortices. This problem is not due to the generation, and roll-up, of the
boundary layers at the boundaries. Further results by the authors have shown that
SPH can reproduce the correct boundary layer roll-up that is seen in
experimental results \cite{wells07vortices}. Furthermore, Violeau and Issa 
\cite{violeau07numerical} showed that SPH can produce accurate log-laws for the
velocity of turbulence flow near a no-slip boundary. The problem occurs once
these boundary layers are advected into the flow, where they are dissipated
before they can interact with the flow to form long-term coherent vortices. 

This paper has also examined the kinetic energy spectrum of the SPH decaying
turbulence, using a Fourier transform that operates directly on the disordered
particles. The spectrum shows the classical $k^{-3}$ spectrum that is expected
from a direct enstrophy cascade. However, for high wavenumbers the spectrum
unexpectedly flattens, indicating an excess of kinetic energy in the SPH
results at small scales. The excess energy is initially seen length scales less
than 7.5 particle spacings and grows in size as the turbulence decays. 

However, the excess kinetic energy at small scales and the lack of the correct
long-term evolution of these vorticity filaments seems to have little impact on
the integral variables of the decaying turbulence (e.g. total kinetic energy,
average squared wavenumber), all of which are reproduced satisfactorily by the
SPH implementation. Future work in this area will focus on evaluating the
performance of SPH at simulating continually forced turbulence.  In this case,
the dissipative characteristics of the method can be more easily evaluated by
examining the energy balance between the power input via the forcing term, the
viscous dissipation and the inverse energy cascade rate.  The SPH DNS
turbulence results will also be used as a baseline to compare with SPH
turbulence models, most notably the consistent XSPH turbulence model proposed
by Monaghan \cite{monaghan09turbulence}.

\bibliographystyle{wileyj}
\bibliography{global.bib}
\end{document}